\documentclass[final,5p,11pt,twocolumn,authoryear]{elsarticle}

\usepackage{graphicx,amsmath,url,amsfonts,booktabs,nicefrac,multirow,microtype,bm,upgreek,threeparttable,slashbox,dblfloatfix,subcaption,appendix}

\usepackage[mathscr]{eucal}

\def\rot#1{\rotatebox{90}{#1}}

\usepackage{hyperref,cleveref}
\hypersetup{colorlinks=true,linkcolor=blue,urlcolor=blue,citecolor=blue}

\journal{Journal of NeuroImage}

\makeatletter
\def\ps@pprintTitle{%
 \let\@oddhead\@empty
 \let\@evenhead\@empty
 \def\@oddfoot{}%
 \let\@evenfoot\@oddfoot}
\makeatother

\begin{document}

\title{Robust parametric modeling of Alzheimer's disease progression}

\author[add1,add2,add3,add4]{Mostafa Mehdipour Ghazi\corref{cor1}}
\ead{mehdipour@biomediq.com}
\author[add1,add2,add3]{Mads Nielsen}
\author[add1,add2,add3]{Akshay Pai}
\author[add4,add5]{Marc Modat}
\author[add4,add5]{M. Jorge Cardoso}
\author[add4,add5]{S\'ebastien Ourselin}
\author[add1,add2,add3]{Lauge S{\o}rensen}
\author{for the Alzheimer's Disease Neuroimaging Initiative\corref{cor2}}

\cortext[cor1]{Corresponding Author.}
\cortext[cor2]{Data used in the preparation of this article were obtained from the Alzheimer's Disease Neuroimaging Initiative (ADNI) database (adni.loni.usc.edu). As such, the investigators within the ADNI contributed to the design and implementation of ADNI and/or provided data but did not participate in analysis or writing of this report. A complete listing of ADNI investigators can be found at \url{http://adni.loni.usc.edu/wp-content/uploads/how_to_apply/ADNI_Acknowledgement_List.pdf}}
\address[add1]{Biomediq A/S, Copenhagen, DK}
\address[add2]{Cerebriu A/S, Copenhagen, DK}
\address[add3]{Department of Computer Science, University of Copenhagen, Copenhagen, DK}
\address[add4]{Department of Medical Physics and Biomedical Engineering, University College London, London, UK}
\address[add5]{School of Biomedical Engineering and Imaging Sciences, King's College London, London, UK}

\begin{frontmatter}

\begin{abstract}
Quantitative characterization of disease progression using longitudinal data can provide long-term predictions for the pathological stages of individuals. This work studies the robust modeling of Alzheimer's disease progression using parametric methods. The proposed method linearly maps the individual's age to a disease progression score (DPS) and jointly fits constrained generalized logistic functions to the longitudinal dynamics of biomarkers as functions of the DPS using M-estimation. Robustness of the estimates is quantified using bootstrapping via Monte Carlo resampling, and the estimated inflection points of the fitted functions are used to temporally order the modeled biomarkers in the disease course. Kernel density estimation is applied to the obtained DPSs for clinical status classification using a Bayesian classifier. Different M-estimators and logistic functions, including a novel type proposed in this study, called modified Stannard, are evaluated on the data from the Alzheimer's Disease Neuroimaging Initiative (ADNI) for robust modeling of volumetric magnetic resonance imaging (MRI) and positron emission tomography (PET) biomarkers, cerebrospinal fluid (CSF) measurements, as well as cognitive tests. The results show that the modified Stannard function fitted using the logistic loss achieves the best modeling performance with an average normalized mean absolute error (NMAE) of $0.991$ across all biomarkers and bootstraps. Applied to the ADNI test set, this model achieves a multiclass area under the ROC curve (AUC) of $0.934$ in clinical status classification. The obtained results for the proposed model outperform almost all state-of-the-art results in predicting biomarker values and classifying clinical status. Finally, the experiments show that the proposed model, trained using abundant ADNI data, generalizes well to data from the National Alzheimer's Coordinating Center (NACC) with an average NMAE of $1.182$ and a multiclass AUC of $0.929$.
\end{abstract}

\begin{keyword}
Alzheimer's disease, disease progression modeling, M-estimation, generalized logistic function, kernel density estimation, Bayesian classifier, magnetic resonance imaging, positron emission tomography, cerebrospinal fluid.
\end{keyword}

\end{frontmatter}

\section{Introduction}

\begin{figure*}[!t]
\centering
\includegraphics[scale=0.8]{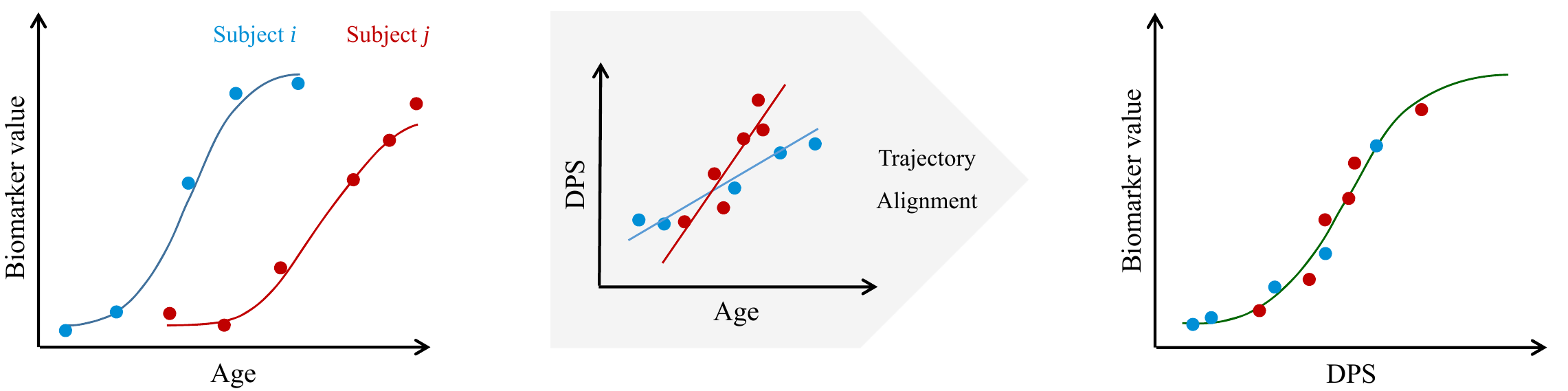}
\caption{An illustration of the AD progression modeling method proposed by \citet{Jedynak2012,Jedynak2015}. Left: A Sigmoid function is fitted to the biomarker measurements of each subject. Middle: The biomarker trajectories are aligned by linearly transforming subject age to DPS. Right: The aligned biomarker fit is obtained for all subjects.}
\label{figure00}
\end{figure*}

Alzheimer's disease (AD) is the most common type of dementia and leads to progressive neurodegeneration, affecting memory and behavior according to regional damage to the brain cells \citep{Ewers2011}. The hippocampus, which is the center of learning and memory, is often one of the first regions of the brain to be damaged. It has also been shown that cerebrospinal fluid (CSF) biomarkers can become abnormal in the presymptomatic phase of the disease, preceding positron emission tomography (PET) and magnetic resonance imaging (MRI) biomarkers followed by clinical markers \citep{Jack2010,Biagioni2011}. Currently, the cause of AD is not clear, and there is no cure or effective treatment to stop its progression, but early diagnosis of the disease, especially in the pre-symptomatic stages, can provide time to treat symptoms and plan for the future. However, early diagnosis of AD is challenging mainly because elderly subjects can suffer from different age-related pathologies and normal aging besides AD. Therefore, methods to stage and identify at-risk individuals are critical to dementia research.

Disease progression modeling uses longitudinal studies to develop data-driven quantitative models that describe the evolution of the disease over time. The approach can, therefore, provide a complete perspective of the disease by computationally exploring the available data to help with a better understanding of the disease for diagnostic, staging, monitorization, and prognostic purposes. Parametric disease progression modeling methods can be divided into two categories, continuous fitting for modeling the dynamics of biomarkers and discrete ordering for biomarker abnormality detection, both relying on unsupervised learning, e.g., by using maximum-likelihood estimation. The discrete methods focus on temporally ordering of biomarkers becoming abnormal during the disease stages by discretizing the disease progression trajectory using generative, event-based models \citep{Fonteijn2012,Venkatraghavan2019}.

Continuous parametric methods for modeling the progression of Alzheimer's disease have been inspired by hypothetical models assuming a sigmoidal evolution of AD biomarkers \citep{Jack2010,Jack2013}. The goal of these methods is to model biomarker values as a function of disease progression \citep{Jedynak2012, Oxtoby2017}. Accordingly, a variety of approaches have been applied to fit a continuous function to the longitudinal dynamics of each biomarker using statistical models such as differential equations and mixed-effects models \citep{Oxtoby2014,Yau2015,Guerrero2016,Bilgel2016,Li2019}, in which one needs to align the trajectory of individuals based on some time measure, e.g., time-to-conversion. These methods are simple and require less data, but parametric assumptions on the biomarker trajectories limit the flexibility of the fits.

Recently, nonparametric disease progression modeling methods have been introduced to model biomarkers jointly while taking temporal dependencies among measurements into account using Gaussian processes \citep{Lorenzi2017} or deep learning \citep{Ghazi2019}. These methods are flexible, make fewer assumptions for modeling, and do not require alignment of the trajectories of the individuals. However, a multivariate gaussian process with monotonicity constraints is computationally expensive to fit due to large covariance matrices, and deep learning methods are less interpretable, hard to train in cases when the data is sparse or irregular, and cannot easily be applied for prediction when the unseen data has fewer biomarkers than what was used for training.

\begin{figure*}[!t]
\centering
\begin{subfigure}[t]{0.34\textwidth}
\raisebox{-\height}{\includegraphics[scale=0.57]{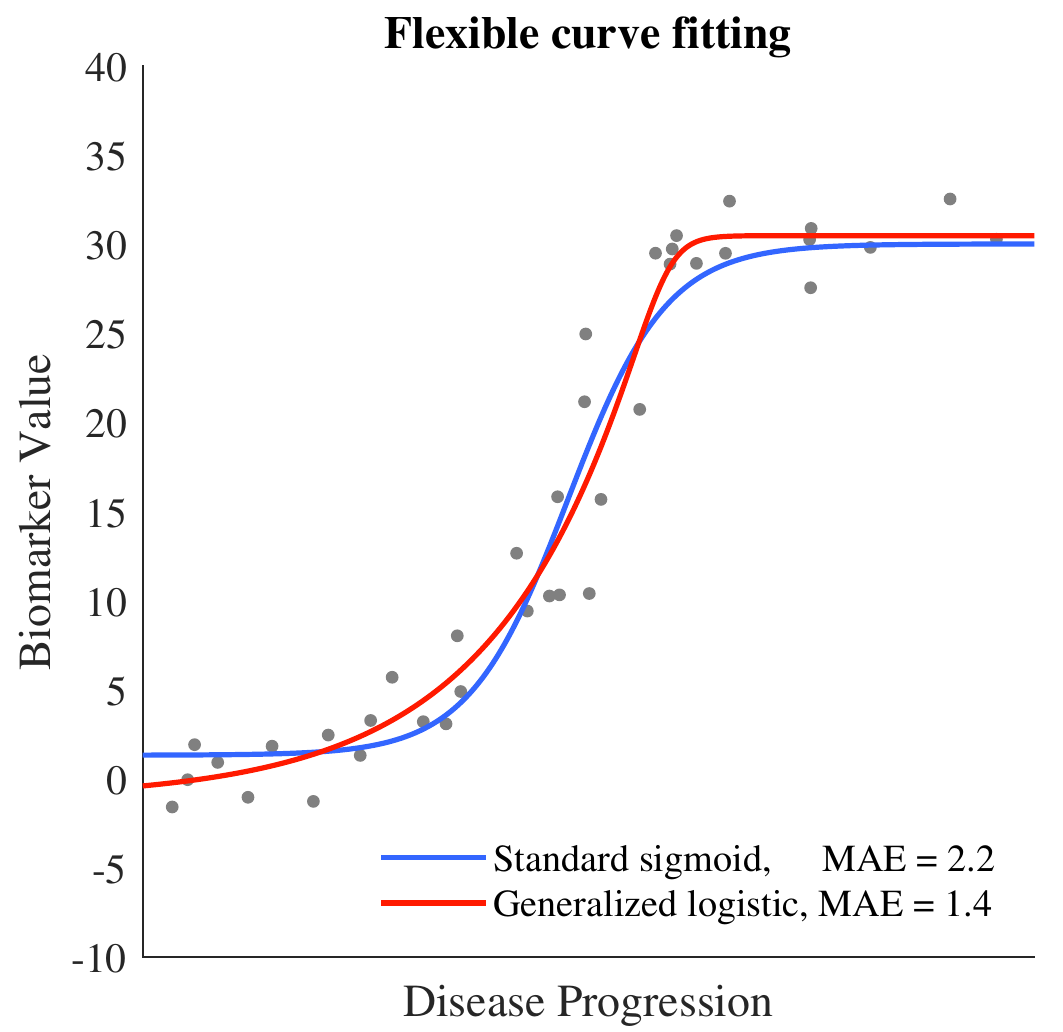}}
\end{subfigure}
\begin{subfigure}[t]{0.32\textwidth}
\raisebox{-\height}{\includegraphics[scale=0.57]{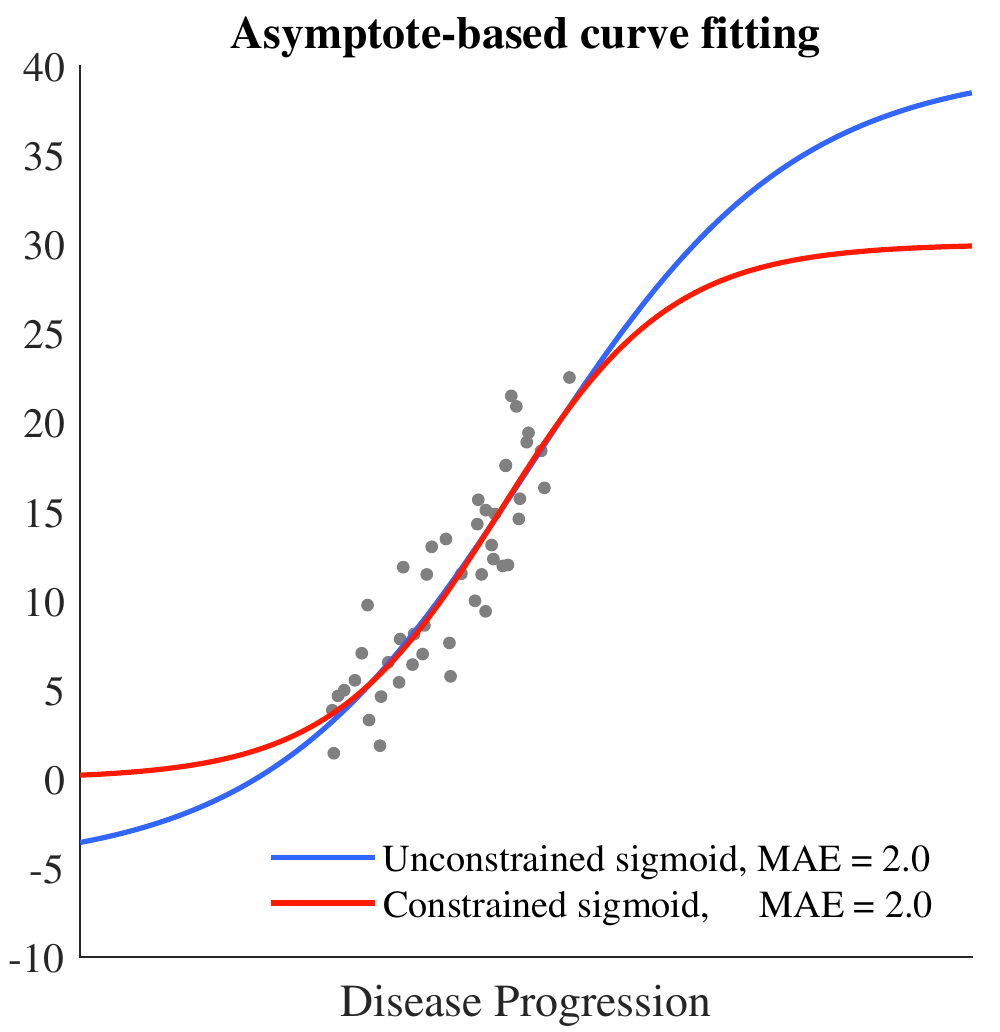}}
\end{subfigure}
\begin{subfigure}[t]{0.32\textwidth}
\raisebox{-\height}{\includegraphics[scale=0.57]{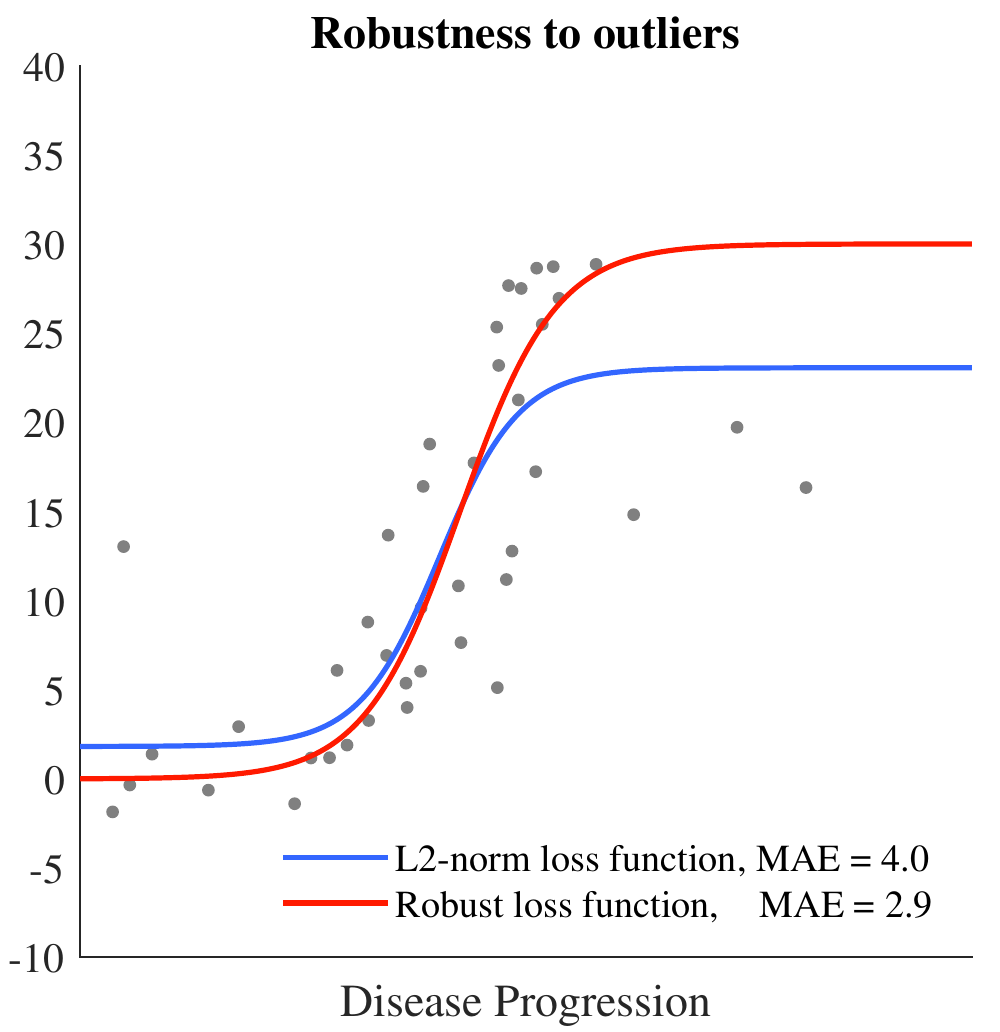}}
\end{subfigure}
\caption{An illustration of how the proposed method (red curves) tackles the existing biomarker curve-fitting problems using simulated data generated based on logistic functions and additive white Gaussian noise. Left: A flexible function is used to fit the asymmetric shape of the simulated data points. Middle: A constrained function is utilized to estimate the exact dynamics of the biomarker. Right: A robust estimator is applied to fit a curve to the simulated data contaminated with outliers.}
\label{figure0}
\end{figure*}

\citet{Jedynak2012,Jedynak2015} proposed a parametric algorithm, which requires less time-series data to train and incorporates information from multiple biomarkers for modeling progression of AD over a common disease timescale. As shown in Figure \ref{figure00}, the method linearly transforms the age of the individual to a disease progression score (DPS) for the time-wise alignment of within-cohort measurements, assuming that the visit intervals in the data are short relative to the disease duration. Alternating least squares is applied to fit a sigmoid function to the longitudinal dynamics of each biomarker. In this method, biomarker trajectories are fitted independently and the biomarker dependencies are only considered when the algorithm alternates to estimate the subject-specific (age) parameters, which in turn can cause difficulties for the convergence of the alternating algorithm. Furthermore, the proposed model is not robust to outliers that can be found in more contaminated data. The first two problems have been tackled by \citet{Bilgel2019}, but the problem with outliers remains.

Maximum likelihood type estimation (M-estimation) is introduced as a robust regression method \citep{Huber2004} to cope with outliers by minimizing a loss function designed to de-emphasize outliers (see Table \ref{table2}). The model fit can further be improved by utilizing a more flexible function (see Table \ref{table1}) and/or constraining the objective function. However, increasing the number of parameters needs to be penalized as it can increase the model complexity and result in overfitting. Figure \ref{figure0} illustrates 1) how the use of a flexible function improves the curve fit, 2) how the use of a constrained function moves lower and upper asymptotes to fit the exact dynamics of the biomarker, and 3) how the use of M-estimation reduces the influence of outliers.

In this paper, a robust extension of \citet{Jedynak2012,Jedynak2015} is proposed that jointly fits a constrained logistic function to the longitudinal dynamics of each biomarker using M-estimation to address the abovementioned curve-fitting problems, e.g., outliers, in the biomarker modeling (see Figure \ref{figure0}). The estimated parameters are quantified using bootstrapping via Monte Carlo resampling, and the inflection points are used to temporally order the biomarkers in the disease course. Kernel density estimation with normal bases is applied to the estimated DPSs for clinical status classification using a Bayesian classifier. Different loss and logistic functions are considered, including a modified version of the Stannard function \citep{Stannard1985} that tends to better describe the biomarker trajectories, and they are applied to AD progression modeling of the Alzheimer's Disease Neuroimaging Initiative (ADNI) \citep{Petersen2010} and the National Alzheimer’s Coordinating Center (NACC) \citep{Beekly2007} data using volumetric MRI biomarkers, CSF and PET measurements, and cognitive tests. The obtained results indicate that the modified Stannard function fitted using the logistic loss achieves the best modeling performance over different bootstraps, and it consistently outperforms the basic algorithm of \citet{Jedynak2015} and state-of-the-art results of \citet{Bilgel2019} and \citet{Marinescu2020} in almost all experiments.

The main contribution of this study is four-fold. First, we introduce a novel generalized logistic function, called modified Stannard, which better fits the AD biomarker trajectories compared to using other logistic functions. Second, the use of M-estimation suppresses the effect of outliers on the fit. Third, the across-cohort generalizability of the proposed model is evaluated by applying the model trained using ADNI data to the test data from the NACC cohort with fewer biomarkers. Finally, an end-to-end approach is introduced that performs biomarker trajectory modeling (unsupervised learning), biomarker inflection point detection (event ordering), and clinical status classification (supervised learning). This is a holistic way to implement a system suitable for both research and clinical applications to better study, detect, and monitor AD.

\section{The proposed method}

The main objective of this work is to minimize the error of parametric disease progression modeling while making the estimates stable and robust to outliers. This is achieved by fitting a constrained logistic function using an M-estimator.

\begin{table*}[!t]
\centering
\small
\caption{Details of the utilized logistic functions for AD progression modeling. Note that the range of each function can be controlled by two additional parameters.}
\label{table1}
\renewcommand{\arraystretch}{1.75}
\centering
\begin{tabular}{l|l|c|c|l|c}
\toprule
\multicolumn{1}{l}{Logistic function} & \multicolumn{1}{c}{$g(s;\bm{\theta})$} & \multicolumn{1}{c}{$\bm{\theta}$} & \multicolumn{1}{c}{$(min, max) \, \forall \, b > 0$} & \multicolumn{1}{c}{$g'(s;\bm{\theta})$} & \multicolumn{1}{c}{$g''(c;\bm{\theta})$} \\
\bottomrule
\addlinespace[0.7pt]
Verhulst & $\left[1 + e^{-b(s-c)}\right]^{-1}$ & $\{b,c\}$ & $(0, 1)$ at $(-\infty, +\infty)$ & $be^{-b(s-c)}\left[1 + e^{-b(s-c)}\right]^{-2}$ & 0 \\
Gompertz & $e^{-e^{-b(s-c)}}$ & $\{b,c\}$ & $(0, 1)$ at $(-\infty, +\infty)$ & $be^{-b(s-c)}e^{-e^{-b(s-c)}}$ & 0 \\
Richards & $\left[1 + \gamma e^{-b(s-c)}\right]^{-1 / \gamma}$ & $\{b,c,\gamma\}$ & $(0, 1)$ at $(-\infty, +\infty)$ & $be^{-b(s-c)}\left[1 + \gamma e^{-b(s-c)}\right]^{-1 -1 / \gamma}$ & 0 \\
Modified Stannard & $\left[1 + \frac{1}{\gamma} e^{-\frac{b}{\gamma}(s-c)}\right]^{-\gamma}$ & $\{b,c,\gamma\}$ & $(0, 1)$ at $(-\infty, +\infty)$ & $\frac{b}{\gamma}e^{-\frac{b}{\gamma}(s-c)}\left[1 + \frac{1}{\gamma} e^{-\frac{b}{\gamma}(s-c)}\right]^{-1 -\gamma}$ & 0 \\
\addlinespace[0.7pt]
\toprule
\end{tabular}
\end{table*}

\subsection{Modeling dynamics of biomarkers}

Two sets of parameters are estimated in the model: observed biomarker-specific parameters, which are assigned for fitting the biomarker curves, and hidden subject-specific (age-related) disease progression parameters that are defined for aligning the trajectory of subjects. Assume that $y_{i,j,k}$ is the $k$-th biomarker's value at the $j$-th visit of the $i$-th subject and $f(s;\bm{\theta})$ is an S-shaped logistic function of DPS $s$ with parameters $\bm{\theta}$. Each biomarker measurement is defined as
\begin{flalign*}
& y_{i,j,k} = f(s_{i,j};\bm{\theta}_k) + \sigma_k \epsilon_{i,j,k} \,, &
\end{flalign*}

\noindent where $\sigma_k$ is the standard deviation of the $k$-th biomarker with $\bm{\theta}_k$ parameters, $\epsilon_{i,j,k}$ is additive white Gaussian noise (random effect) with i.i.d. assumption, and $s_{i,j}$ is the DPS for the $j$-th visit of the $i$-th subject and is obtained as
\begin{flalign*}
& s_{i,j} = \alpha_i t_{i,j} + \beta_i \,, &
\end{flalign*}

\noindent where $t_{i,j}$ is the age of subject $i$ in visit $j$, and $\alpha_i \in \mathbb{R}_{> 0}$ and $\beta_i \in \mathbb{R}$ are the rate and onset of disease progression of subject $i$, respectively. Finally, the multiobjective optimization for robust nonlinear regression is defined as
\begin{flalign*}
& \{\hat{\bm{\alpha}}, \hat{\bm{\beta}}, \hat{\mathbf{\uptheta}}\} = \underset{i,j,k}\min \sum\limits_{i,j,k} w_i \rho\left(\frac{y_{i,j,k} -  f(\alpha_i t_{i,j} + \beta_i;\bm{\theta}_k)}{\sigma_k}\right) \,, &
\end{flalign*}

\noindent where $\rho(\cdot)$ is a maximum likelihood-type function and $w_i = 1 / N_i$ is a weighting factor for normalizing the objective function with the number of available points per subject ($N_i$).

For fitting the longitudinal trajectories of biomarkers, four logistic functions are considered (Table \ref{table1}). All functions have the same range $(0, 1)$ and can produce the same inflection points at $c \in \mathbb{R}$, to be later used for biomarker ordering. We candidate utilization of a modified flexible logistic function based on the Stannard function \citep{Stannard1985}, where the $1 / \gamma$ factor is multiplied by the exponential term to create an asymmetric growth curve with an inflection point at $c$ like other functions. This function tends to better describe asymmetrical sigmoid patterns of the biomarker trajectories with modeling both slow and rapid growths at the beginning or the end of the disease period. In the defined functions, $b \in \mathbb{R}_{> 0}$ and $\gamma \in \mathbb{R}_{> 0}$ denote the growth rate and symmetry parameter of the curves, respectively. The reason for restricting $b$ to the positive real numbers is to make parameters of the estimation identifiable.

It can also be deduced from Table \ref{table1} that the sigmoid function first introduced by Verhulst \citep{Verhulst1845} is a special (symmetric) case of both Richards' function \citep{Richards1959} and the proposed function when $\gamma = 1$. Moreover, Gompertz's function \citep{Gompertz1825} is a simplified form of Richards' function when $\gamma$ approaches zero, i.e., $\lim_{\gamma \to 0} (1 + \gamma u)^{-1 / \gamma} = e^{-u}$. Finally, the upper and lower asymptotes of the curves can be adjusted by two additional parameters \citep{Zwietering1990} as
\begin{flalign*}
& f(s;\bm{\theta}) = (a - d) g(s;\bm{\theta}) + d \,. &
\end{flalign*}

The range parameters, $a$ and $d$, can be set to fixed values when the exact range of biomarkers is given, which is the case with cognitive tests. This, in turn, not only reduces the number of optimization parameters but also increases the stability of the estimation. For other biomarkers, if there are, for example, sign constraints which are the cases with nonnegative CSF and PET measurements, both parameters can be constrained to lower and/or upper bounds, but otherwise remain unconstrained.

\subsection{Model fitting}

Alternating approach, as an efficient optimization technique, is applied to solve the problem where the algorithm iteratively estimates $\mathbf{\uptheta}$ using fixed values of $\bm{\alpha}$ and $\bm{\beta}$ and vice versa until the parameters converge. The proposed algorithm can be summarized as follows

\begin{enumerate}
\setlength{\itemindent}{-15pt}
\item[] \textbf{Initialization:} initialize $\{\bm{\alpha}^{(0)}, \bm{\beta}^{(0)}, \mathbf{\uptheta}^{(0)}\}$ using measurements.
\item[] \textbf{Optimization:} iterate $l$ until convergence.

\textbf{Biomarker fitting:} estimate the biomarker-specific parameters using values of all subjects and visits.
\begin{flalign}\label{marker-specific-eq}
& \resizebox{0.78\hsize}{!}{$\hat{\mathbf{\uptheta}}^{(l)} = \underset{i,j,k}\min \sum\limits_{i,j,k} w_i \rho\left(\frac{y_{i,j,k} -  f(\hat{\alpha}_i^{(l-1)} t_{i,j} + \hat{\beta}_i^{(l-1)};\bm{\theta}_k)}{\sigma_k}\right)$} \,, &
\end{flalign}

\textbf{Age mapping:} estimate the subject-specific parameters using values of all biomarkers and visits.
\begin{flalign}\label{subject-specific-eq}
& \resizebox{0.78\hsize}{!}{$\{\hat{\alpha}_i^{(l)}, \hat{\beta}_i^{(l)}\} = \underset{(j,k) \in N_i}\min \sum\limits_{j,k} w_i \rho\left(\frac{y_{i,j,k} -  f(\alpha_i t_{i,j} + \beta_i;\hat{\bm{\theta}}_k^{(l)})}{\sigma_k}\right)$} \,, &
\end{flalign}
\end{enumerate}

\noindent where $N_i$ corresponds to the number of measurements among all biomarkers and visits available for the $i$-th subject. This way, in contrast to \citet{Jedynak2012,Jedynak2015}, biomarkers are fitted jointly. The degrees of freedom of the fit is equal to $\sum_k (N_k - |\bm{\theta}_k|) - 2I$, where $N_k$ is the number of measurements among all subjects and visits available for the $k$-th biomarker, $|\bm{\theta}_k|$ denotes the number of biomarker-specific parameters for the $k$-th biomarker, and $I$ is the number of subjects. Therefore, the algorithm can be applied in case the data contains more than $\sum_k |\bm{\theta}_k| + 2I$ points, and if any subject has at least two distinct points considering all biomarkers and visits.

The utilized maximum likelihood-type functions for robust regression \citep{Holland1977,Pennacchi2008} are described in Table \ref{table2}. These estimators attempt to diminish the influence of the outliers while fitting curves. In general, M-estimators use a tuning parameter called $\tau$ to scale the functions as $\tau^2 \rho(r / \tau)$ in order to yield $95\%$ asymptotic efficiency with respect to the standard normal distribution. The corresponding tuning constants for the utilized functions are also reported in Table \ref{table2}.

\begin{table}[t]
\centering
\small
\caption{The utilized $\rho$-type M-estimators and their corresponding scale factors $\tau$ for robust regression.}
\label{table2}
\renewcommand{\arraystretch}{1.75}
\centering
\begin{tabular}{l|l|c}
\toprule
\multicolumn{1}{l}{Loss function} & \multicolumn{1}{c}{$\rho(r)$} & \multicolumn{1}{c}{$\tau$} \\
\bottomrule
L2 & $r^2$ & 1 \\
L1-L2 & $2\left(\sqrt{\rule{0pt}{2ex} 1 + r^2} - 1\right)$ & 1 \\
Logistic & $\ln(\cosh(r))$ & 1.205 \\
Modified Huber & $\begin{cases} 1 - \cos(\mathopen| r \mathclose|), & \mathopen| r \mathclose| \leq{\pi/2} \\  \mathopen| r \mathclose| + (1 - \pi/2), &  \mathopen| r \mathclose| > \pi/2 \end{cases}$ & 1.2107 \\
Cauchy-Lorentz & $\ln(1 + r^2)$ & 2.3849 \\
\toprule
\end{tabular}
\end{table}

Finally, the obtained DPSs are standardized with respect to the scores of the available cognitively normal visits of subjects in order to calibrate all biomarker trajectories in different experiments. This process removes the mean of the normal visits' distribution of DPSs and scales the range to give a better intuition of timing of disease progression in the course of AD. In this case, it would be necessary to properly update the parameters as
\begin{flalign*}
& s_{i,j} = \left(s_{i,j} - \mu_{cn}\right) / \sigma_{cn} \,, & \\
& \alpha_i = \alpha_i / \sigma_{cn} \,, & \\
& \beta_i = \left(\beta_i - \mu_{cn}\right) / \sigma_{cn} \,, & \\
& b_k = \sigma_{cn} b_k \,, & \\
& c_k = \left(c_k - \mu_{cn}\right) / \sigma_{cn} \,, &
\end{flalign*}

\noindent where $\mu_{cn}$ and $\sigma_{cn}$ are the mean and standard deviation of the DPSs in the available cognitively normal visits of subjects, respectively.

\subsection{Biomarker value prediction}

Biomarker values can be predicted using the fitted model parameters. Age mapping part of the proposed algorithm is applied to estimate the subject-specific parameters using Equation (\ref{subject-specific-eq}) based on the values of those biomarkers of the test subject that have available estimated biomarker-specific parameters in the fitted model. Next, biomarker values are predicted as $f(s_{i,j};\bm{\theta}_k)$ using the estimated test DPSs where $f(\cdot)$ is the logistic function applied to model fitting.

\subsection{Clinical status classification}

In order to predict the clinical status of test subjects per visit, kernel density estimation (KDE) \citep{Parzen1962} is used to fit the likelihoods of cognitively normal, cognitively impaired, and AD groups in a nonparametric way. Assume that $(s_1, s_2, \ldots, s_N)$ is a set of $N$ i.i.d. DPSs sampled from an unknown distribution with density function $p(s | c_i)$, where $c_i$ denotes the $i$-th class label. KDE is expressed as
\begin{flalign*}
& \hat{p}(s | c_i) = \frac{1}{Nw} \sum\limits_{n = 1}^{N} \mathscr{K}\left( \frac{s - s_n}{w} \right) \,, &
\end{flalign*}

\noindent where $\mathscr{K(\boldsymbol{\cdot})}$ is a smooth (kernel) function with a smoothing bandwidth $w > 0$. Here, the Gaussian kernel is used as the smoothing function. 

The clinical status is classified based on the DPSs with a Bayesian classifier that uses the KDE-based fitted likelihoods as
\begin{flalign*}
& p(c_i | s) = \frac{p(c_i) p(s | c_i)}{\sum\limits_i p(c_i) p(s | c_i)} \,, &
\end{flalign*}

\noindent where $p(c_i)$ is the data-driven prior probability for the $i$-th class, $p(c_i | s)$ is the posterior probability for predicting the test DPS that belongs to the class $c_i$, and the term in the denominator specifies the evidence and can be dropped because the maximum a posteriori estimation is used for classification.

\section{Experimental setup}

\subsection{Data}

The data used in this work is obtained from the ADNI and the NACC databases.

\subsubsection{ADNI}

The ADNI was launched in 2003 as a public-private partnership, led by principal investigator Michael W. Weiner, MD. The primary goal of ADNI has been to test whether serial MRI, PET, other biological markers, and clinical and neuropsychological assessment can be combined to measure the progression of mild cognitive impairment and early Alzheimer's disease. We use The Alzheimer's Disease Prediction Of Longitudinal Evolution (TADPOLE) challenge dataset \citep{Marinescu2018} that includes the three ADNI phases ADNI 1, ADNI GO, and ADNI 2. This dataset contains measurements from brain MRI, PET, CSF, cognitive tests, and demographics, and genetic information.

The labels cognitively normal (CN), significant memory concern (SMC), and normal (NL) are merged under CN; mild cognitive impairment (MCI), early MCI (EMCI), and late MCI (LMCI) under MCI; and AD and dementia under AD. In addition, subjects converting from one clinical status to another, e.g., MCI-to-AD, are assigned the latter label (AD in this example). The utilized ADNI data includes T1-weighted brain MRI volumes of ventricles, hippocampus, whole brain, fusiform, and entorhinal cortex, PET scan measures of fludeoxyglucose (FDG-PET) and florbetapir (AV45-PET), CSF measures of Amyloid beta, total tau, and phosphorylated tau, as well as the cognitive tests of clinical dementia rating sum of boxes (CDR-SB), Alzheimer's disease assessment scale 13 items (ADAS-13), mini-mental state examination (MMSE), functional activities questionnaire (FAQ), Montreal cognitive assessment (MOCA), and Rey auditory verbal learning test of immediate recall (RAVLT-IR). Detailed information about the utilized biomarkers can be found in Table \ref{table_list}.

\subsubsection{NACC}

The NACC, established by the National Institute on Aging of the National Institutes of Health in 1999, has been developing a large database of standardized clinical and neuropathological data from both exploratory and explanatory Alzheimer's disease research \citep{Beekly2007}. The data has been collected from different Alzheimer's disease centers across the United States and among others contains measurements from different modalities such as MRI, PET, and cognitive tests.

Labels with numerical cognitive status of one (normal cognition) and two (impaired-not-MCI) are merged under CN, and cognitive status of three (MCI) and four (Dementia) are set to MCI and AD, respectively. It should be noted that we only keep subjects with primary etiologic diagnosis of normal, AD, or missing. This is to exclude subjects diagnosed with other types of dementia, non-neurodegenerative disease, or a non-neurological condition. The used NACC data includes T1-weighted brain MRI volumes of hippocampus and whole brain, and the cognitive tests of MMSE, MOCA, FAQ (sum of the 10-item scores), and CDR-SB using the CDR{\textregistered} Dementia Staging Instrument.

\begin{table*}[t]
\centering
\normalsize
\begin{threeparttable}
\caption{Demographics of the obtained datasets after filtering across visits.}
\label{table3}
\renewcommand{\arraystretch}{1.25}
\centering
\begin{tabular}{llccc}
\toprule
\multicolumn{2}{c}{clinical status} & age, year & education, year & MMSE \\
 & & (mean$\pm$SD) & (mean$\pm$SD) & (mean$\pm$SD) \\
\bottomrule
 & CN & 76.93$\pm$6.03 & 15.76$\pm$2.92 & 29.05$\pm$1.20 \\
 & MCI & 75.07$\pm$7.67 & 15.80$\pm$2.90 & 27.43$\pm$2.26 \\
 & AD & 76.47$\pm$7.51 & 15.80$\pm$2.90 & 21.61$\pm$4.61 \\
\rot{\rlap{~~~~~ADNI}} & Missing & 74.44$\pm$7.87 & 16.10$\pm$2.64 & 27.34$\pm$3.07 \\
\midrule
 & CN & 79.06$\pm$7.34 & 13.76$\pm$4.00 & 28.46$\pm$1.71 \\
 & MCI & 80.83$\pm$8.57 & 13.79$\pm$4.03 & 25.32$\pm$3.03 \\
 & AD & 81.09$\pm$8.14 & 13.73$\pm$4.08 & 19.60$\pm$5.11 \\
\rot{\rlap{~~~~~NACC}} & Missing & 78.88$\pm$11.69 & 13.56$\pm$4.69 & 28.29$\pm$2.36 \\
\toprule
\end{tabular}
\vspace{-0.05in}
\begin{tablenotes}
\item {\footnotesize Note that missing clinical status after filtering is indicated as \lq{}Missing\rq{}.}
\end{tablenotes}
\vspace{0.05in}
\end{threeparttable}
\end{table*}

\begin{table*}[t]
\centering
\normalsize
\caption{Statistics of the visits per dataset after filtering.}
\label{table4}
\renewcommand{\arraystretch}{1.25}
\centering
\begin{tabular}{lccccccc}
\toprule
 & \multicolumn{4}{c}{\# visits per clinical status} & \# visits per subject & visit interval, year & \# measurements per subject \\
 & CN & MCI & AD & Missing & (mean$\pm$SD) & (mean$\pm$SD) & (mean$\pm$SD) \\
\bottomrule
ADNI & 2,285 & 3,850 & 2,064 & 899 & 5.99$\pm$2.37 & 0.74$\pm$0.43 & 58.60$\pm$23.38 \\
NACC & 1,140 & 205 & 318 & 9 & 7.00$\pm$2.91 & 1.15$\pm$0.37 & 21.61$\pm$9.03 \\
\toprule
\end{tabular}
\end{table*}

\subsubsection{Data filtering}

For our analysis, in each of the ADNI and NACC datasets, measurements outside known biomarker ranges, e.g., RAVLT-IR $< 0$, are rejected and assumed as missing values. The volumetric MRI outliers observed in the ADNI dataset are removed by assuming intracranial volume (ICV) estimates that are proportionally smaller than estimated corresponding MRI measurements, i.e., MRI / ICV $> 1$, as missing values.

Clinical follow-up visits with reverting clinical diagnoses are removed per subject considering the neighboring visits. In the ADNI dataset, clinical follow-up visits with wrongly ordered dates are discarded per subject. Also, MRI, PET, and CSF measurements that are already matched to the cognitive visits with any extreme time gaps are excluded. The acceptable time gap is obtained based on the data statistics and is set to three months. In the NACC dataset, we perform the matching of MRI and clinical visits. However, due to the relatively smaller sample size in NACC compared with ADNI, matches more than three months apart are kept and treated as two distinct visits. In this analysis, we assign a missing clinical status for any MRI visits that do not fall within the 3-month window.

In order to be able to apply the proposed method, measurements and clinical diagnoses with missing date information per visit are set to missing values, and subjects with less than two distinct visits are omitted. This results in 219 ADNI subjects and 151 NACC subjects being excluded.

\subsubsection{The obtained study population}

After filtering the data, the utilized 16 ADNI biomarkers are acquired from 1,518 subjects (854 males and 664 females) in 9,098 visits between August 2005 and May 2017, and the six NACC biomarkers are acquired from 239 subjects (75 males and 164 females) in 1,672 visits between October 2005 and July 2018. All subjects in both datasets have at least one cognitive test. In the NACC data, 203 subjects underwent MRI imaging while in the ADNI data, 1,515 and 1,220 subjects underwent MRI and PET imaging, respectively, and 1,088 subjects have CSF measures. Table \ref{table3} and Table \ref{table4} summarize statistics of the demographics and measurements in the two datasets after data filtering. Note that both datasets include missing values and missing clinical status, the latter indicated as \lq{}Missing\rq{}.

\subsubsection{Data preprocessing}

In the ADNI dataset, the volumetric measurements were obtained using two different versions of the public FreeSurfer software package \citep{Fischl2002}, and in the NACC dataset, they were calculated using IDeA Lab following ADNI protocols. Therefore, the MRI measurements need to be corrected for FreeSurfer version \citep{Gronenschild2012}, software package, and hence for different cohorts (ADNI-NACC). In addition, biological difference in brain size hinders direct utilization of MRI measurements for disease progression estimation. Total intracranial volume (TIV) or ICV is a commonly used measure for normalization to correct for head size. To overcome both aforementioned problems of difference in cohort/software (version) and head size, we employ the residual approach \citep{OBrien2011} based on the analysis of covariance, which takes data from control groups and linearly regresses MRI volumes on their corresponding ICV as a covariate of interest. The corrected measurements can thus be calculated as the estimated residuals $\hat{R}$ of the volumes using the regression parameters obtained from the control data as
\begin{flalign*}
& \hat{R}_{i,j,k,v} = ROI_{i,j,k,v} - \left[\hat{\beta}_{k,v}^{cn} + \hat{\alpha}_{k,v}^{cn} ICV_{i,j,k,v}\right] \,, &
\end{flalign*}

\noindent where $ROI_{i,j,k,v}$ is the $k$-th MRI volume for subject $i$ at visit $j$ calculated (observed) using software or software version $v$, $ICV$ is the corresponding intracranial volume, and $\hat{\beta}^{cn}$ and $\hat{\alpha}^{cn}$ are the intercept and slope of the regression estimated from the CN group. Finally, the estimated residuals are standardized per cohort/software (version) so that all variables have zero mean and unit variance.

\subsubsection{Data partitioning and bootstrapping}

To evaluate the algorithms, each of the ADNI and NACC datasets is partitioned into two non-overlapping sets for training and testing. To be more specific, based on the first and last available diagnoses of subjects, i.e., CN-CN, CN-MCI, ..., AD-AD, we divide each of these types of pairs into two groups including few and many visits using the median number of visits as threshold and randomly select $20\%$ of the subjects from each group for testing.

Additionally, bootstrapping is used in the experiments for quantification of the estimation, and in each bootstrap, a subset of the training subjects is randomly sampled with replacement based on the first and last available pair of diagnoses and the number of available visits per subject, to make sure each bootstrap sampling contains data from any diagnostic status and sequence lengths. The unused subjects are assigned for validation and account for $1 / e \approx 0.37$ of the subjects where $e$ is the base of the natural logarithm. This also means that the estimated variance using the bootstrapped model will account for approximately $63\%$ of the total variance.

To facilitate future research in AD progression modeling and comparison with the current study, all the data splits, including each bootstrap split, are also provided on arXiv as ancillary files.

\subsection{Evaluation metrics}

Robust Bayesian information criterion (BIC) is used as a criterion for model selection among the robust models \citep{Machado1993}. The criterion is penalized with the number of parameters to avoid overfitting, where the model with the lowest BIC is preferred, and it is defined as
\begin{flalign*}
& \mathrm{BIC} = 2 E_{train}^{(L_{opt})} + Q \ln(N)  \,, &
\end{flalign*}

\noindent where $E_{train}^{(L_{opt})}$ is the training loss at the optimum iteration number $L_{opt}$ obtained through biomarker fitting using Equations (\ref{marker-specific-eq}) and (\ref{subject-specific-eq}), $N$ is the total number of measurements, and $Q$ is the total number of parameters which is equal to $\sum_k |\bm{\theta}_k| + 2I$.

The mean absolute error (MAE) is used to assess the modeling performance as a measure less sensitive to outliers \citep{Chai2014}. It is calculated based on the absolute differences between actual and estimated values as follows
\begin{flalign*}
& \mathrm{MAE} = \frac{1}{N_k} \sum\limits_{(i,j)\in{N_k}} \big|y_{i,j,k} - f(s_{i,j};\bm{\theta}_k)\big| \,, &
\end{flalign*}

\noindent where $N_k$ is the number of measurements among all subjects and visits available for the $k$-th biomarker, and $y_{i,j,k}$ and $f(s_{i,j};\bm{\theta}_k)$ are the ground-truth and estimated values of the $k$-th biomarker for the $i$-th subject at the $j$-th visit. Absolute errors of different biomarkers can be normalized with the corresponding standard deviation of the biomarkers and averaged across all normalized biomarkers to obtain a single performance measure called normalized MAE (NMAE). The modeling performance of two different methods is statistically compared using the paired, two-sided Wilcoxon signed-rank test \citep{Wilcoxon1945} applied to the NMAEs obtained from different bootstraps.

Additionally, multiclass area under the receiver operating characteristic (ROC) curve (AUC) \citep{Hand2001} is used to measure the diagnostic performance in a multiclass test set and is calculated using the posterior probabilities as
\begin{flalign*}
& \resizebox{0.98\hsize}{!}{$\mathrm{AUC} = \frac{1}{(n_c(n_c-1))} \sum\limits_{i=1}^{n_c-1}\sum\limits_{k=i+1}^{n_c}\frac{1}{n_i n_k} \left[\mathrm{SR}_i - \frac{n_i(n_i+1)}{2}+\mathrm{SR}_k-\frac{n_k(n_k+1)}{2}\right]$} \,, &
\end{flalign*}

\noindent where $n_c$ is the number of distinct classes, $n_i$ denotes the number of available observations belonging to the $i$-th class, and $\mathrm{SR}_i$ is the sum of the ranks of posteriors $p(c_i|\bm{s}_i)$ after sorting all concatenated posteriors $\{p(c_i|\bm{s}_i),p(c_i|\bm{s}_k)\}$ in an ascending order, where $\bm{s}_i$ and $\bm{s}_k$ are vectors of DPSs belonging to the true classes $c_i$ and $c_k$, respectively. Likewise, $\mathrm{SR}_k$ is the sum of the ranks of posteriors $p(c_k|\bm{s}_k)$ after sorting all concatenated posteriors $\{p(c_k|\bm{s}_k),p(c_k|\bm{s}_i)\}$ in an ascending order.

\begin{table*}[!b]
\centering
\normalsize
\begin{threeparttable}
\caption{Modeling performance as BIC (mean$\pm$SD) $\times 10^4$ for the 100-times bootstrapped ADNI training subsets using different logistic and loss functions.}
\label{table5}
\renewcommand{\arraystretch}{1.5}
\centering
\begin{tabular}{l|ccccc}
\toprule
\backslashbox[1pt][lr]{\hspace{-0.08in} Logistic \\ \hspace{-0.08in} function}{Loss function} & L2 & L1-L2 & Logistic & Modified Huber & Cauchy-Lorentz \\
\bottomrule
Verhulst & 2.090$\pm$0.039 & 1.901$\pm$0.028 & 1.830$\pm$0.027 & 1.836$\pm$0.027 & 1.925$\pm$0.029 \\
Gompertz & 2.101$\pm$0.042 & 1.902$\pm$0.028 & 1.831$\pm$0.027 & 1.836$\pm$0.027 & 1.927$\pm$0.029 \\
Richards & 2.077$\pm$0.038 & 1.899$\pm$0.028 & 1.829$\pm$0.027 & 1.835$\pm$0.027 & 1.924$\pm$0.029 \\
Modified Stannard & 2.077$\pm$0.038 & 1.898$\pm$0.028 & \textbf{1.828}$\pm$\textbf{0.026} & 1.834$\pm$0.027 & 1.924$\pm$0.028 \\
\toprule
\end{tabular}
\vspace{-0.05in}
\begin{tablenotes}
\item {\footnotesize The best result is shown in boldface and its corresponding configuration is selected for the remaining experiments.}
\end{tablenotes}
\vspace{0.05in}
\end{threeparttable}
\end{table*}

\subsection{Initialization of the algorithm and optimization}

Since the fitting algorithm is iteratively performed using an alternating approach starting from values optimized in the previous step, initialization is an important step for efficiently reaching the optimum. We initially set $\bm{\alpha^{(0)}}$ and $\bm{\beta^{(0)}}$ to $\bm{1}$ and $\bm{0}$, respectively. Moreover, we initialize the slope of the trajectories ($\bm\lambda$) to either $-1$ or $1$ depending on the diagnoses. A positive slope is considered when the average of the $k$-th biomarker's values for cognitively normal visits is less than that for AD visits and vice versa.

Next, the parameters of the logistic functions are initially estimated as $\gamma_k = 1$, $c_k = 0$, and $b_k = 4 \lambda_k / (a_k - d_k)$, where $d_k$ and $a_k$ are the minimum and maximum of the $k$-th biomarker's values, respectively, provided that the slope $\lambda_k$ is positive, and vice versa if the slope is negative. Finally, we repeat the alternating procedure using the logistic functions and the trust-region algorithm \citep{Coleman1996} considering robust estimators for at most $30$ iterations.

\subsection{Stopping criteria}

To avoid overfitting, the optimal parameter values are selected according to the optimum generalization loss obtained using the following criteria \citep{Prechelt1998}
\begin{flalign*}
& \{\hat{\bm{\alpha}}, \hat{\bm{\beta}}, \hat{\mathbf{\uptheta}}\} = \min_{L_{min} \leq l \leq L_{max}} E_{valid}^{(l)} \,, &
\end{flalign*}

\noindent where $E_{valid}^{(l)}$ is the validation loss at the $l$-th iteration obtained through biomarker fitting using Equations (\ref{marker-specific-eq}) and (\ref{subject-specific-eq}). The minimum number of iterations, $L_{min}$, is set to $10$ to allow for enough training progress. The maximum number of iterations, $L_{max}$, is set to $50$. This avoids unnecessary computations since it was empirically observed that $E_{valid}$ attained a minimum well within this iteration range in all cases.

\section{Results and discussion}

\subsection{Biomarker modeling}

First, the proposed method is applied to model the dynamics of the ADNI biomarkers. Table \ref{table5} illustrates the modeling performance (BIC) for ADNI training subsets obtained from 100 bootstraps using different logistic and loss functions. The combination of the modified Stannard function for biomarker fitting and the logistic loss achieve the best modeling performance with both the lowest average BIC and the smallest standard deviation and a validation NMAE of $0.985\pm0.029$. This configuration will be used in all the remaining experiments.

\begin{figure*}[!t]
\centering
\begin{subfigure}[t]{0.221\textwidth}
\raisebox{-\height}{\includegraphics[scale=0.4]{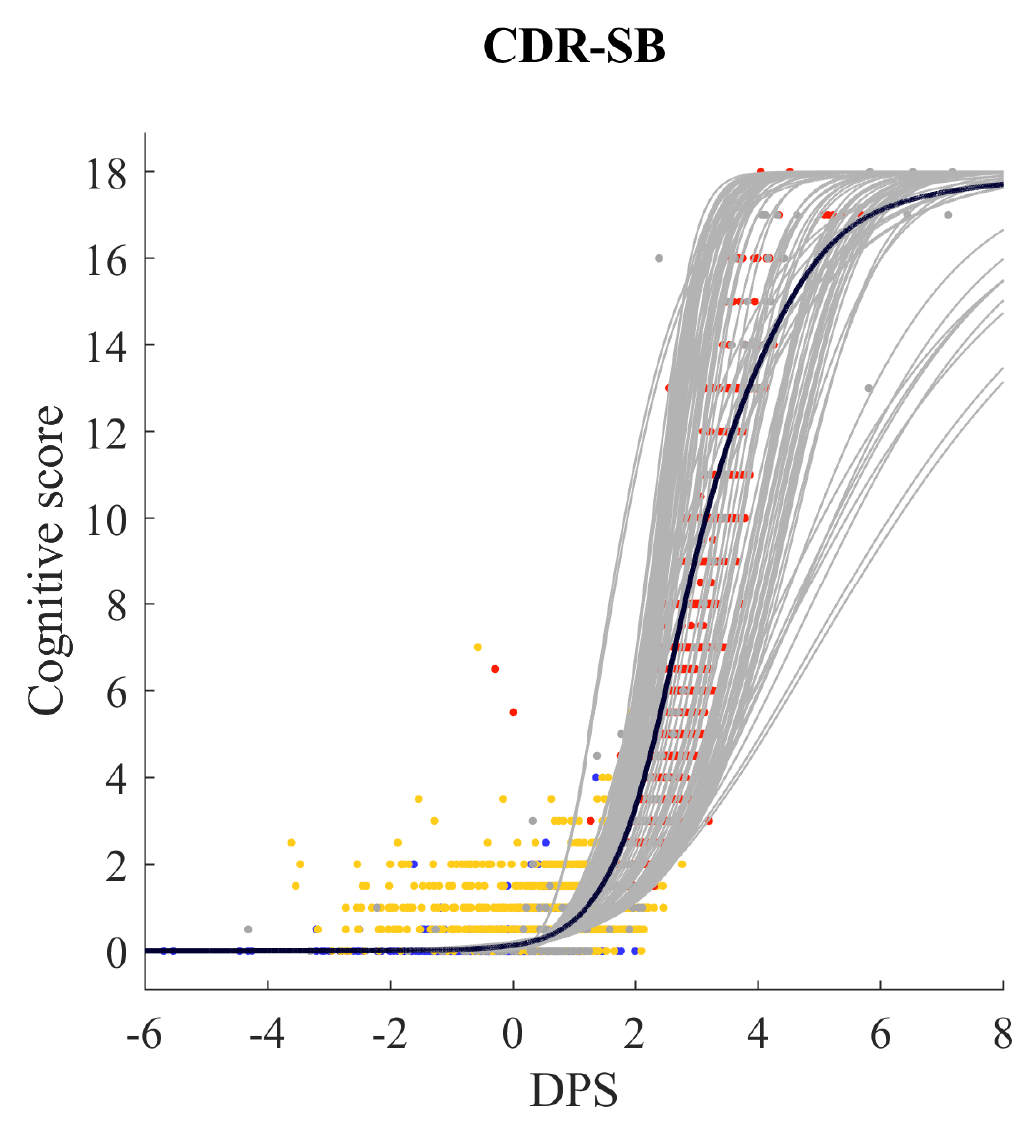}}
\end{subfigure}
\begin{subfigure}[t]{0.221\textwidth}
\raisebox{-\height}{\includegraphics[scale=0.4]{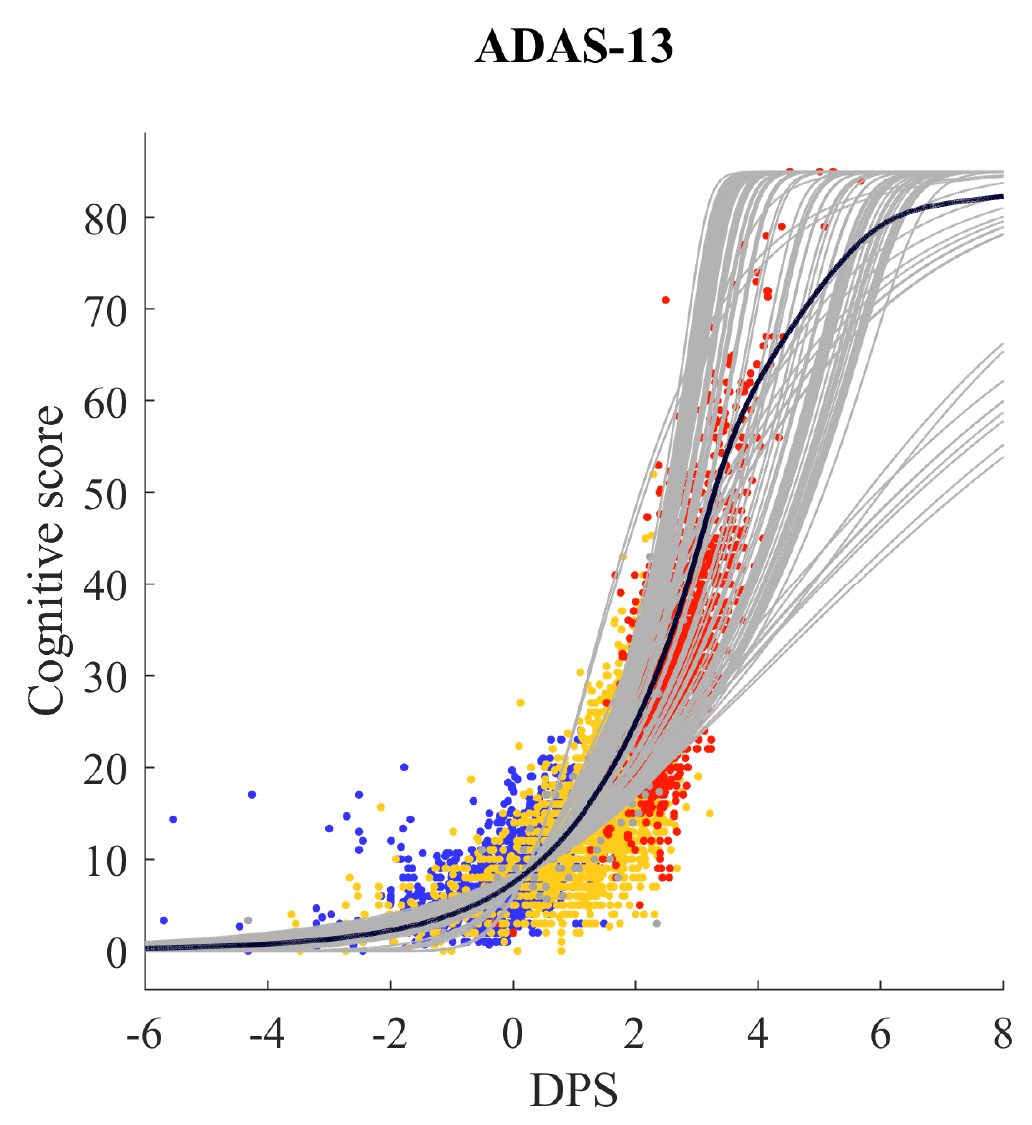}}
\end{subfigure}
\begin{subfigure}[t]{0.221\textwidth}
\raisebox{-\height}{\includegraphics[scale=0.4]{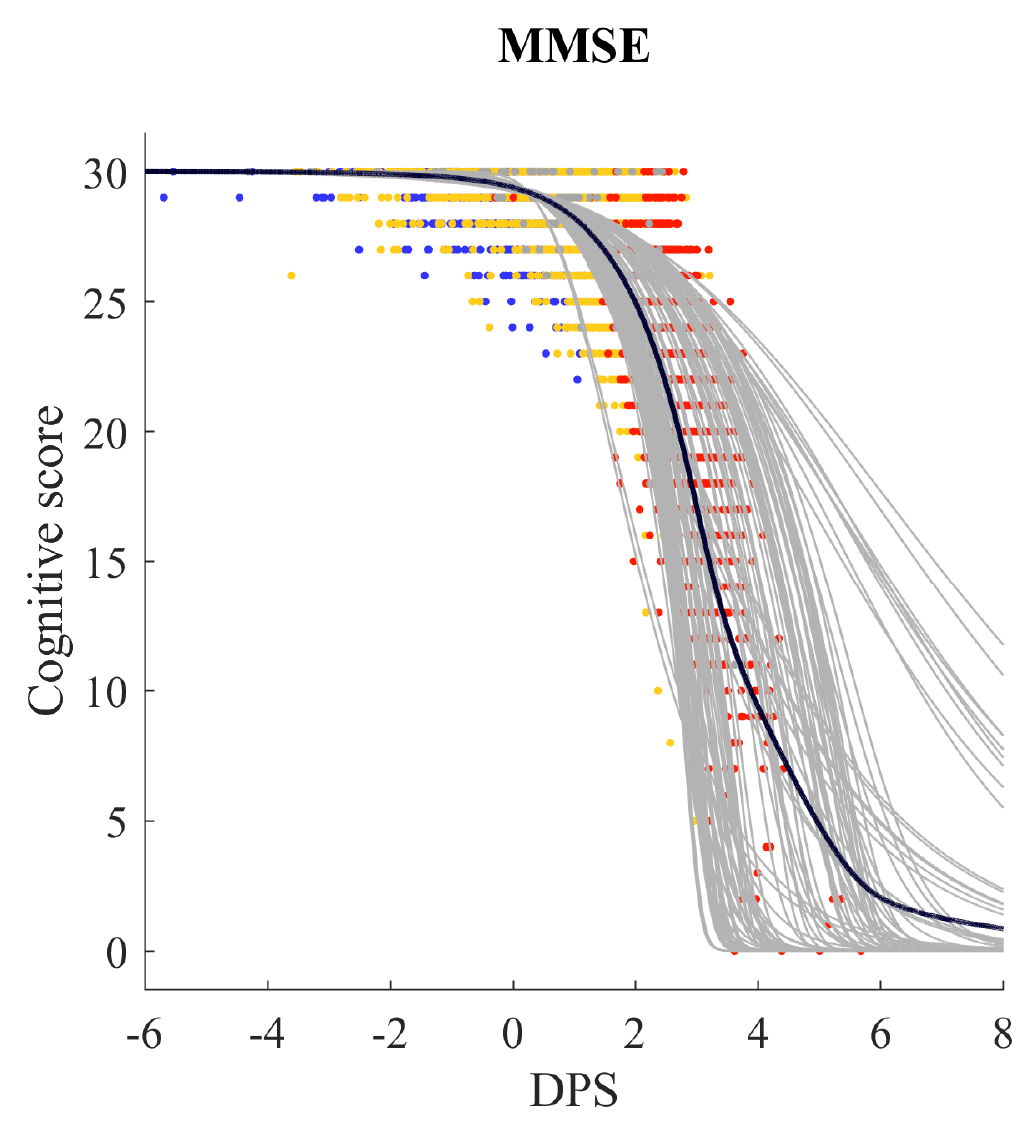}}
\end{subfigure}
\begin{subfigure}[t]{0.321\textwidth}
\raisebox{-\height}{\includegraphics[scale=0.4]{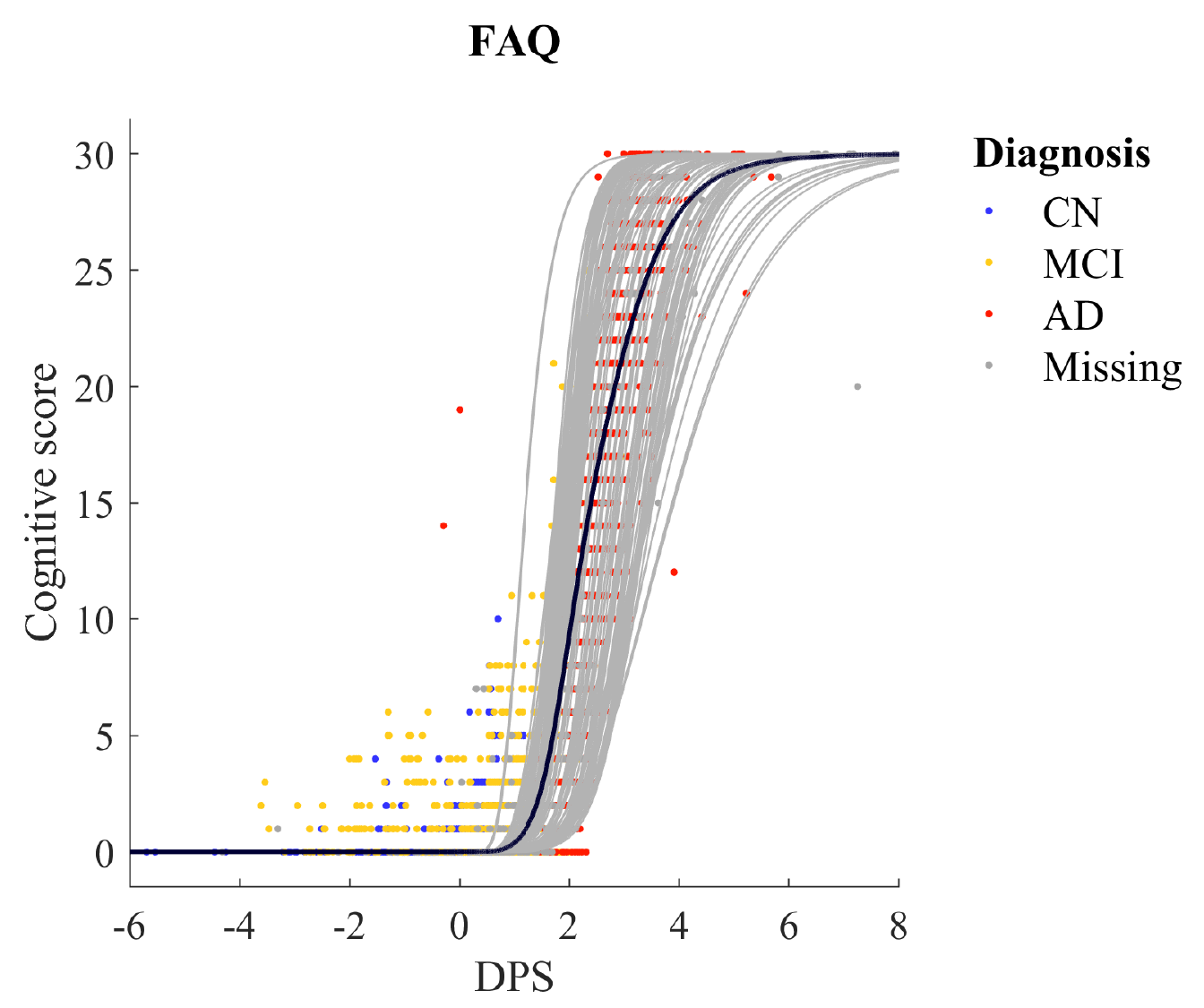}}
\end{subfigure}
\begin{subfigure}[t]{0.221\textwidth}
\raisebox{-\height}{\includegraphics[scale=0.4]{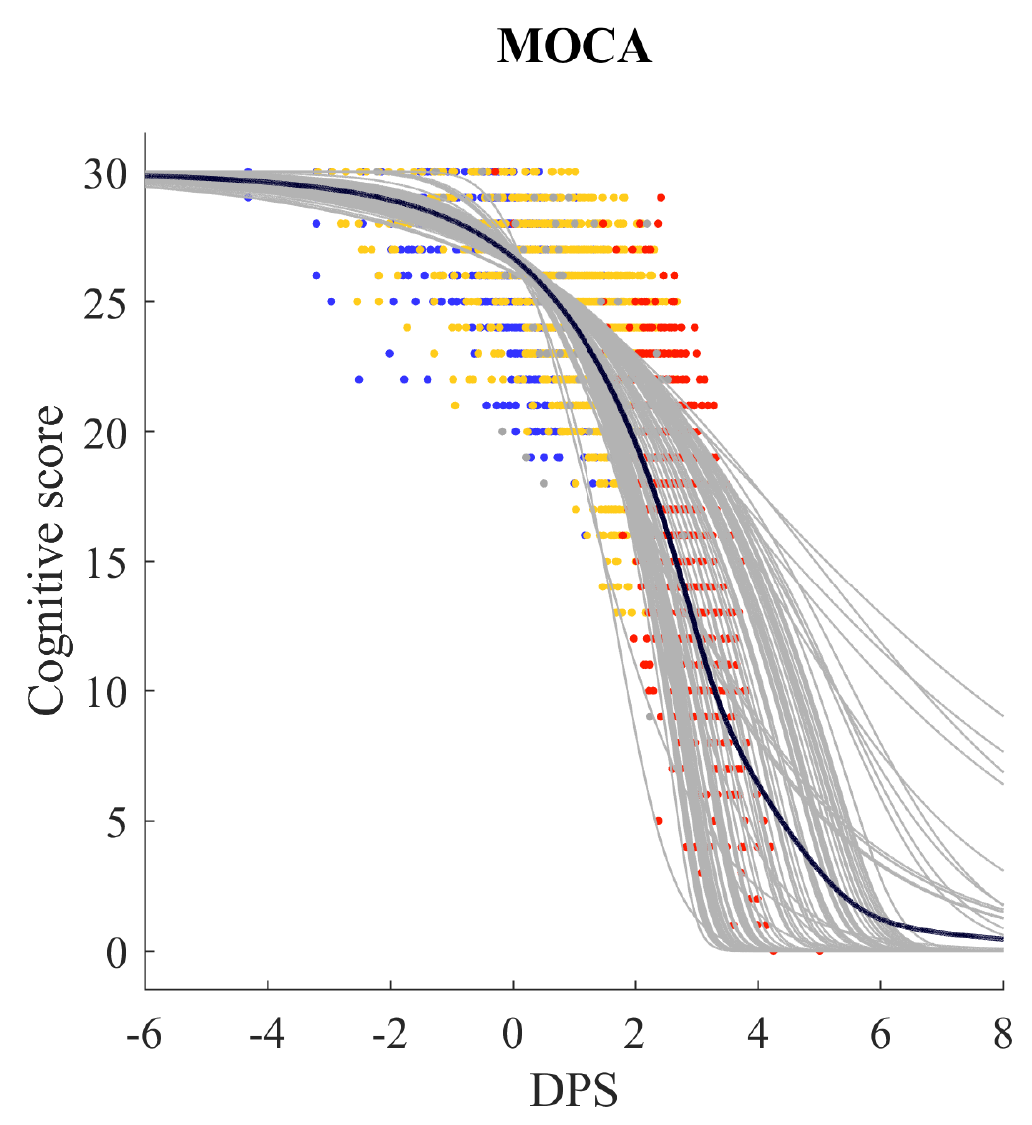}}
\end{subfigure}
\begin{subfigure}[t]{0.221\textwidth}
\raisebox{-\height}{\includegraphics[scale=0.4]{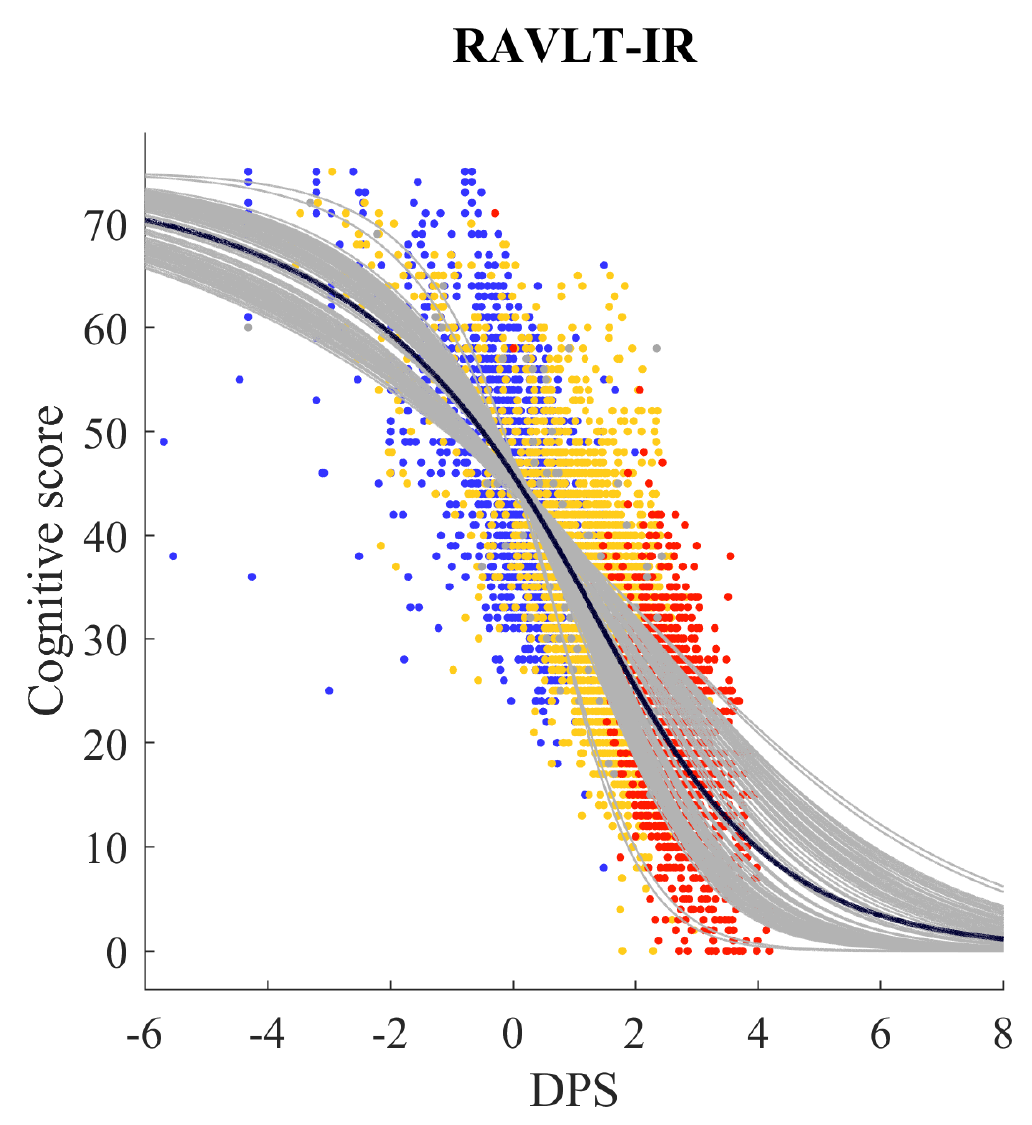}}
\end{subfigure}
\begin{subfigure}[t]{0.221\textwidth}
\raisebox{-\height}{\includegraphics[scale=0.4]{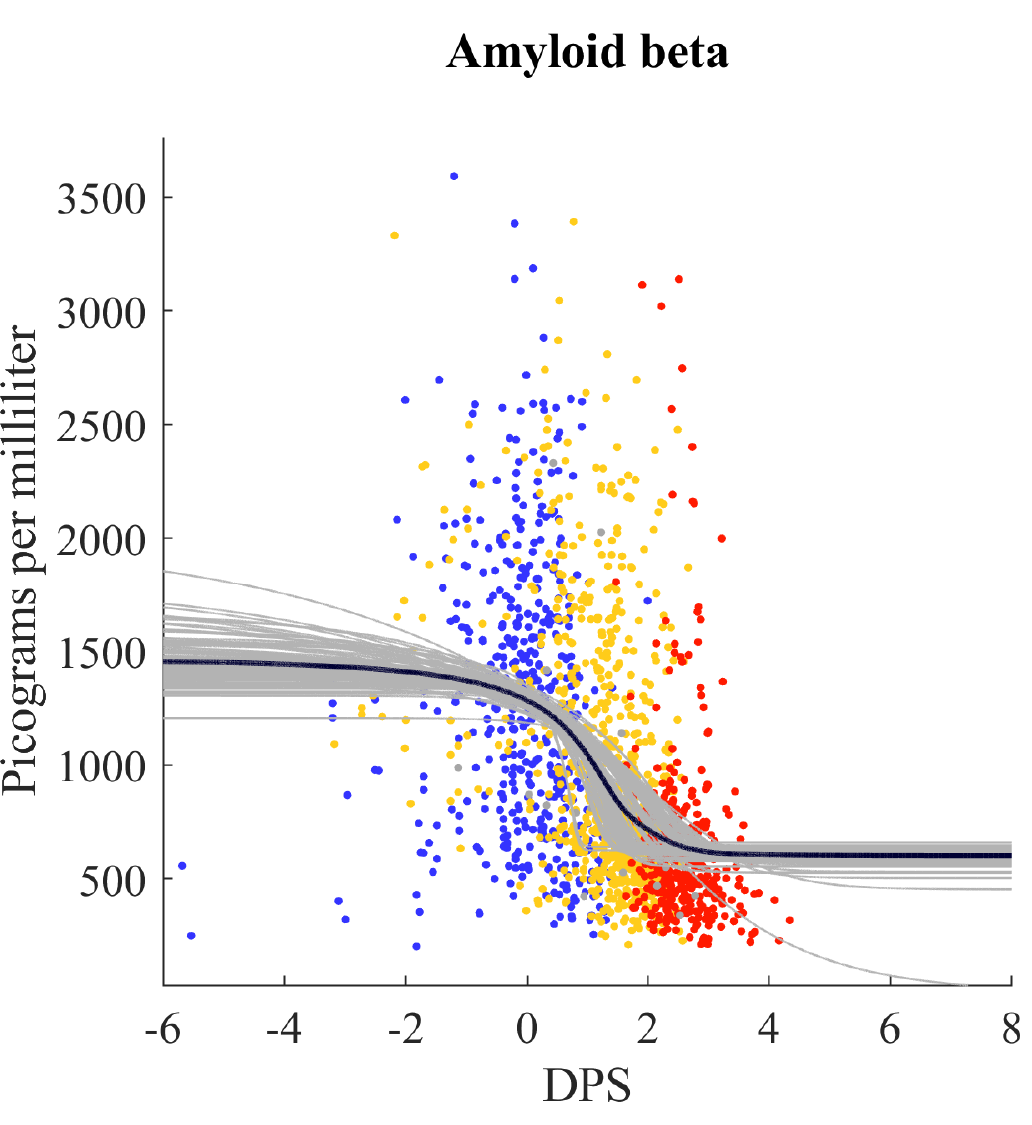}}
\end{subfigure}
\begin{subfigure}[t]{0.321\textwidth}
\raisebox{-\height}{\includegraphics[scale=0.4]{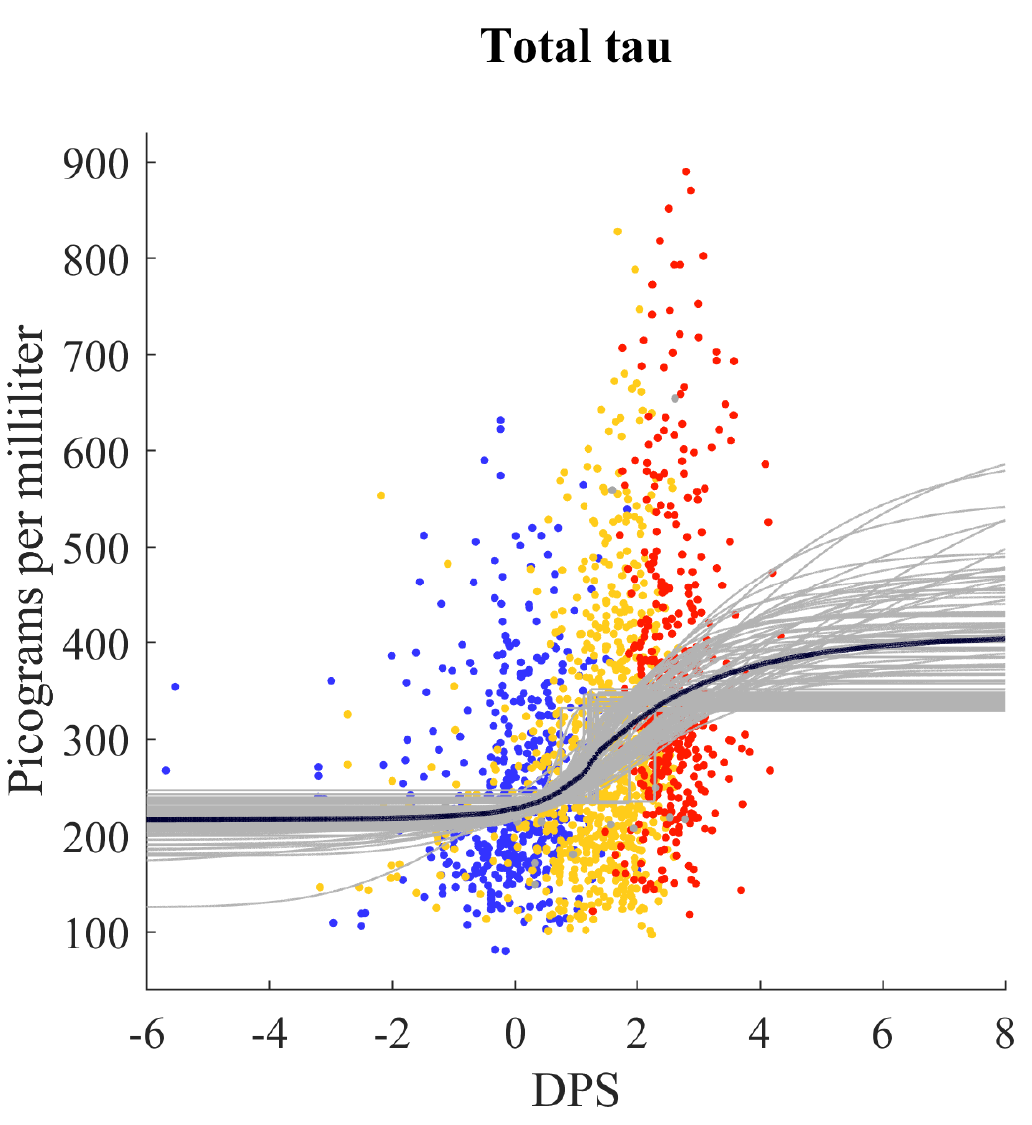}}
\end{subfigure}
\begin{subfigure}[t]{0.221\textwidth}
\raisebox{-\height}{\includegraphics[scale=0.4]{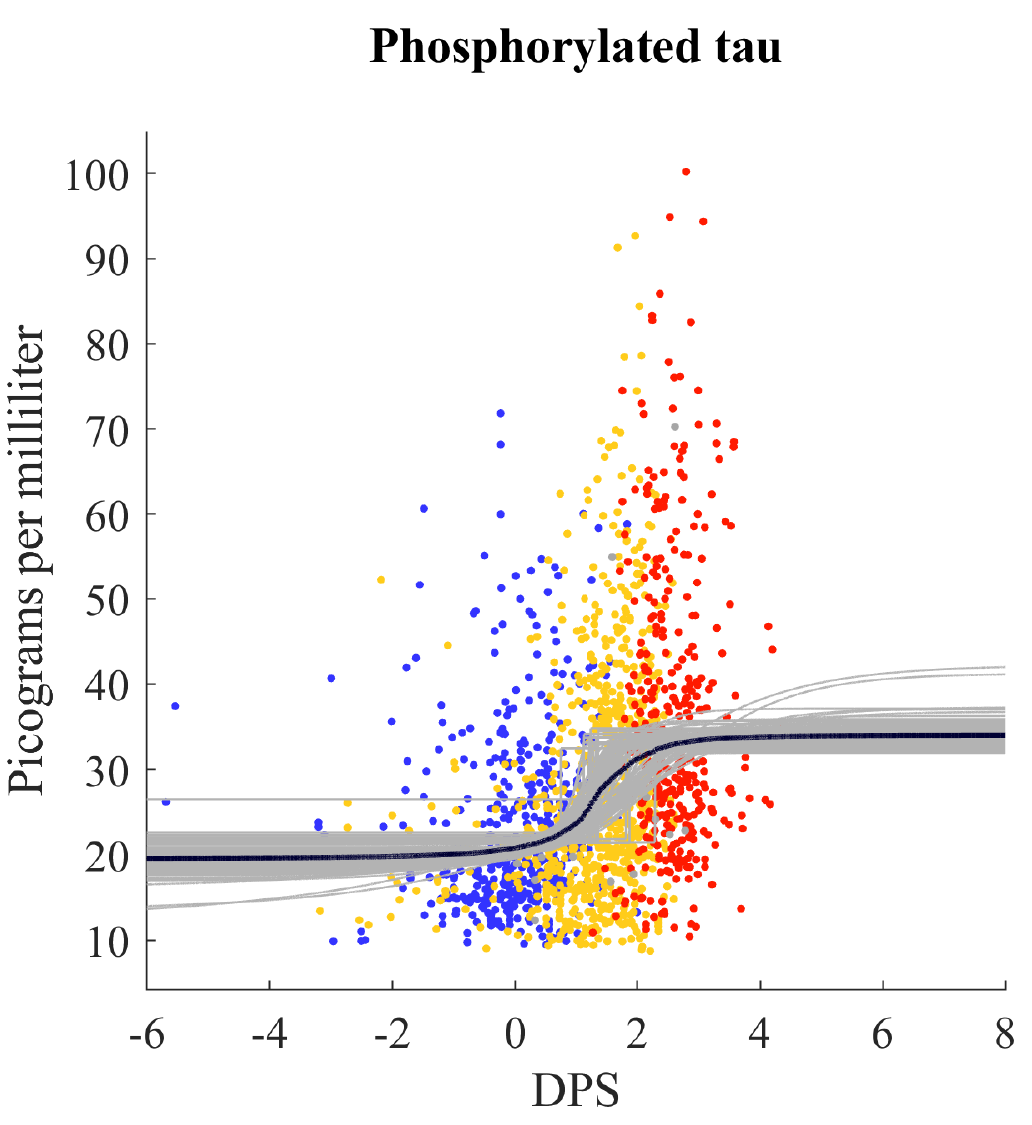}}
\end{subfigure}
\begin{subfigure}[t]{0.221\textwidth}
\raisebox{-\height}{\includegraphics[scale=0.4]{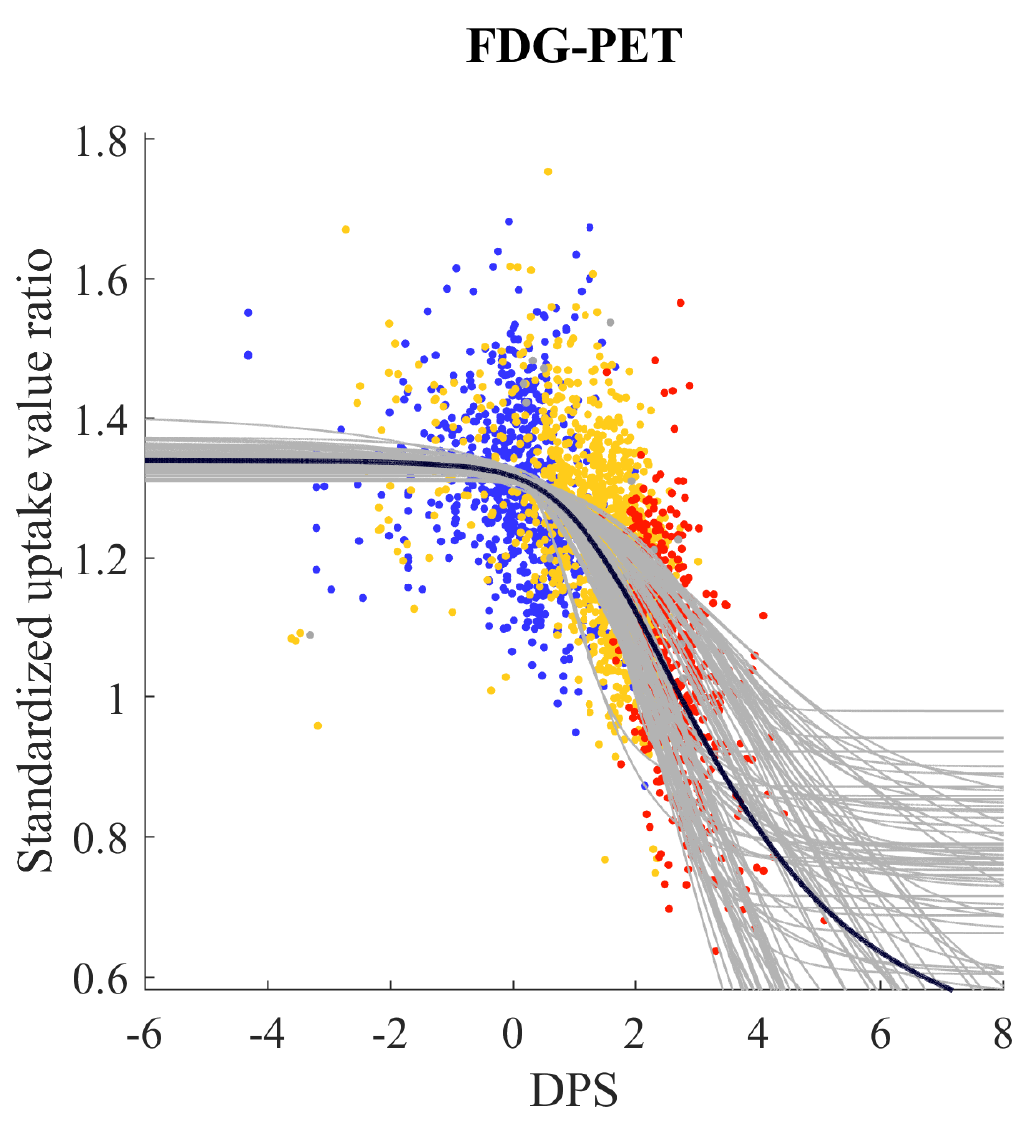}}
\end{subfigure}
\begin{subfigure}[t]{0.221\textwidth}
\raisebox{-\height}{\includegraphics[scale=0.4]{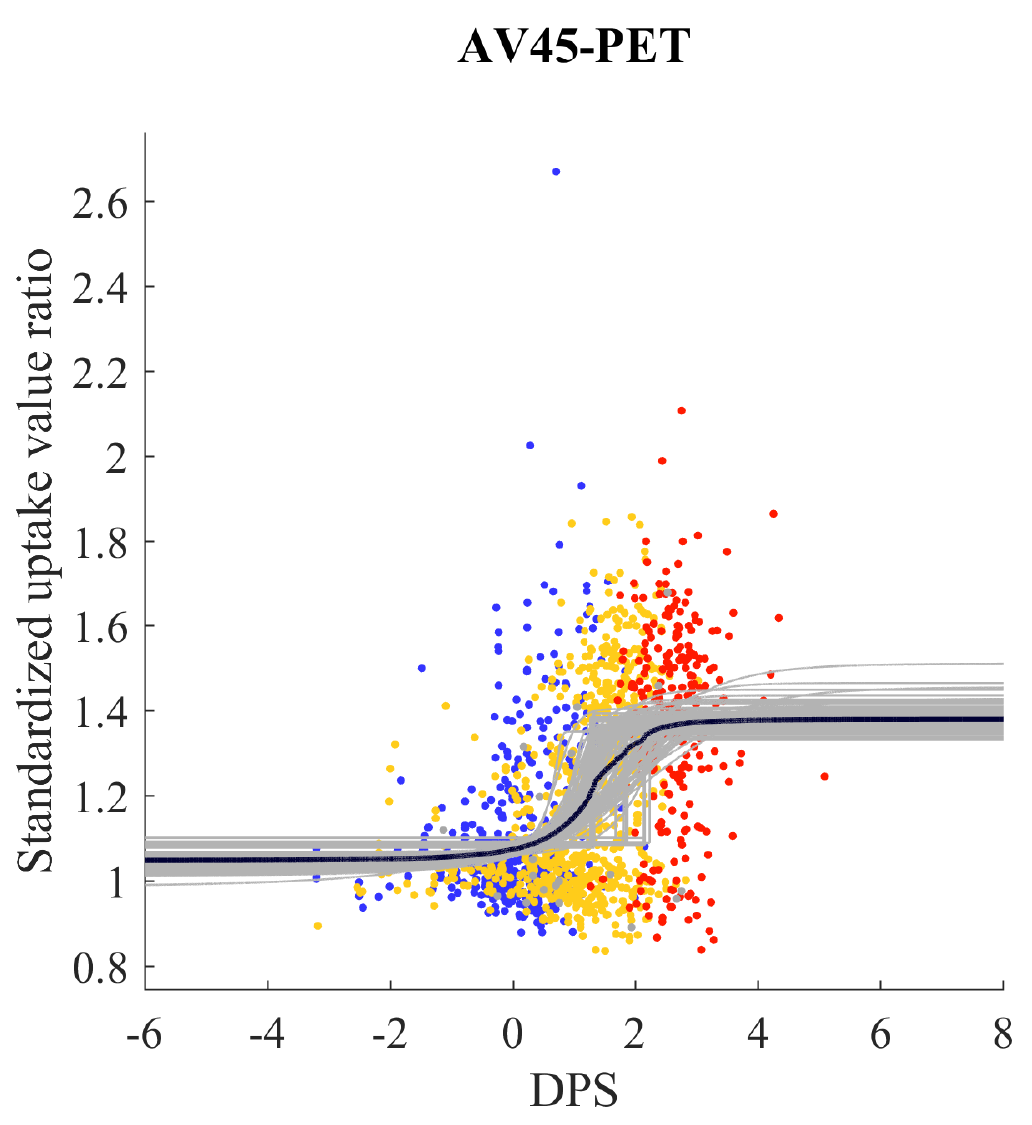}}
\end{subfigure}
\begin{subfigure}[t]{0.321\textwidth}
\raisebox{-\height}{\includegraphics[scale=0.4]{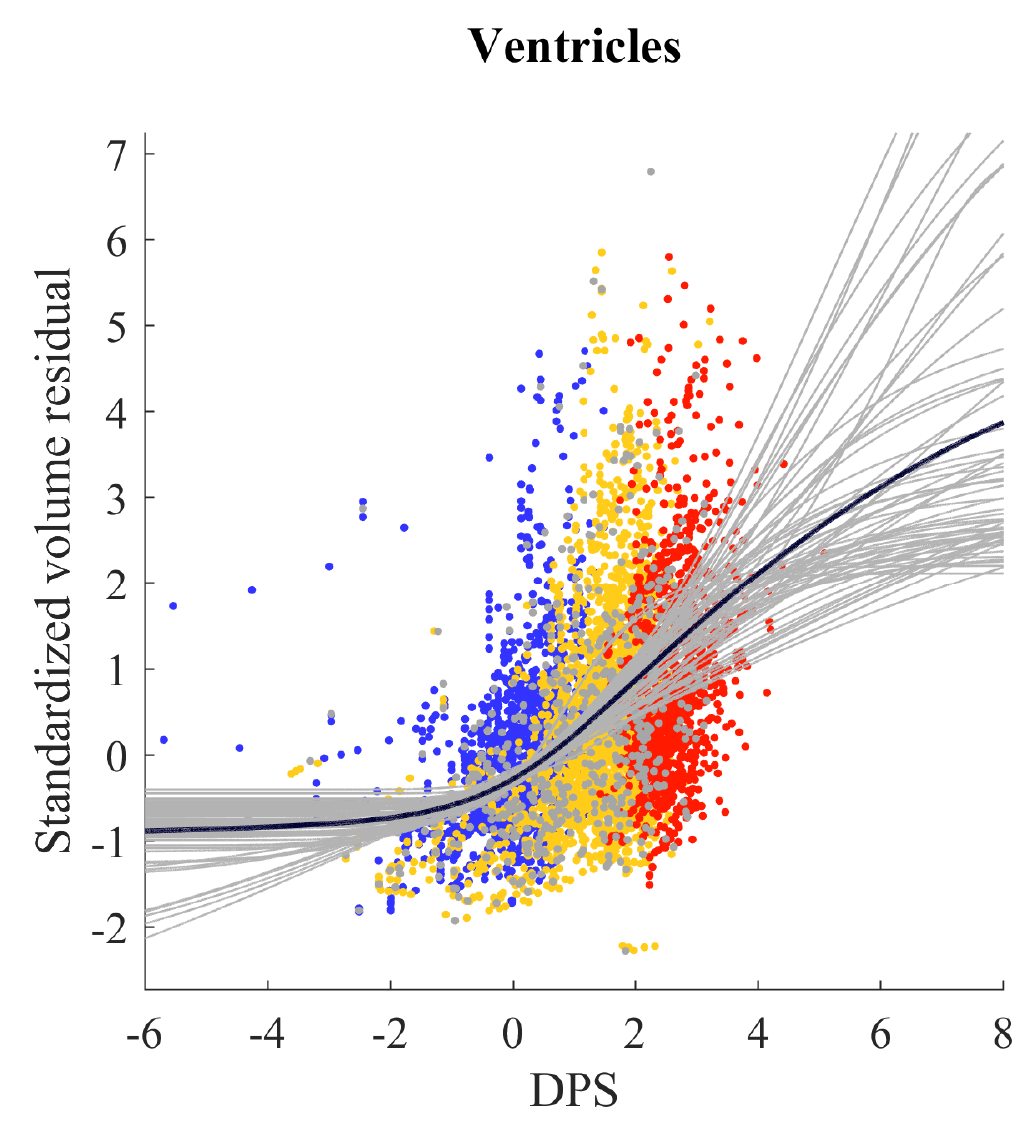}}
\end{subfigure}
\begin{subfigure}[t]{0.221\textwidth}
\raisebox{-\height}{\includegraphics[scale=0.4]{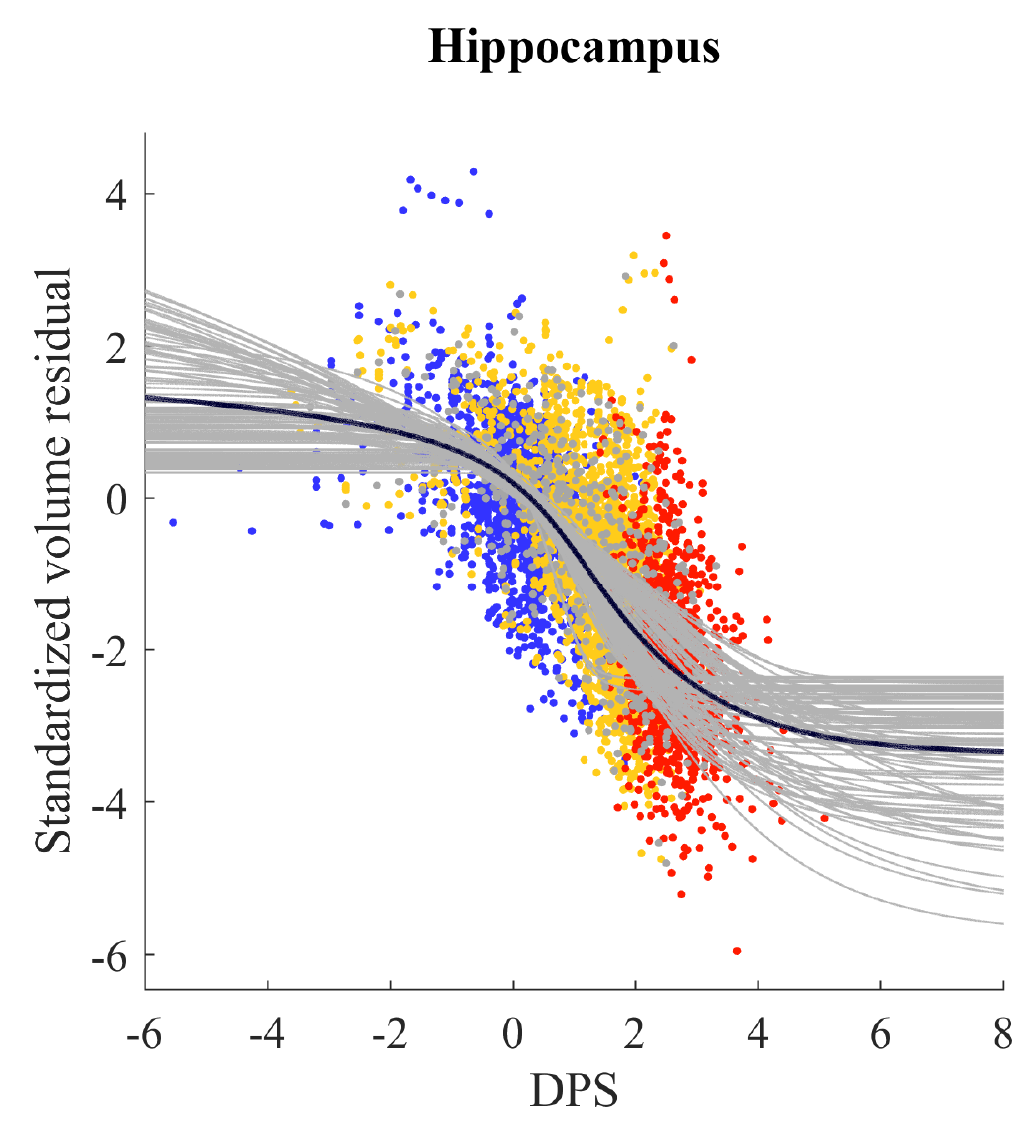}}
\end{subfigure}
\begin{subfigure}[t]{0.221\textwidth}
\raisebox{-\height}{\includegraphics[scale=0.4]{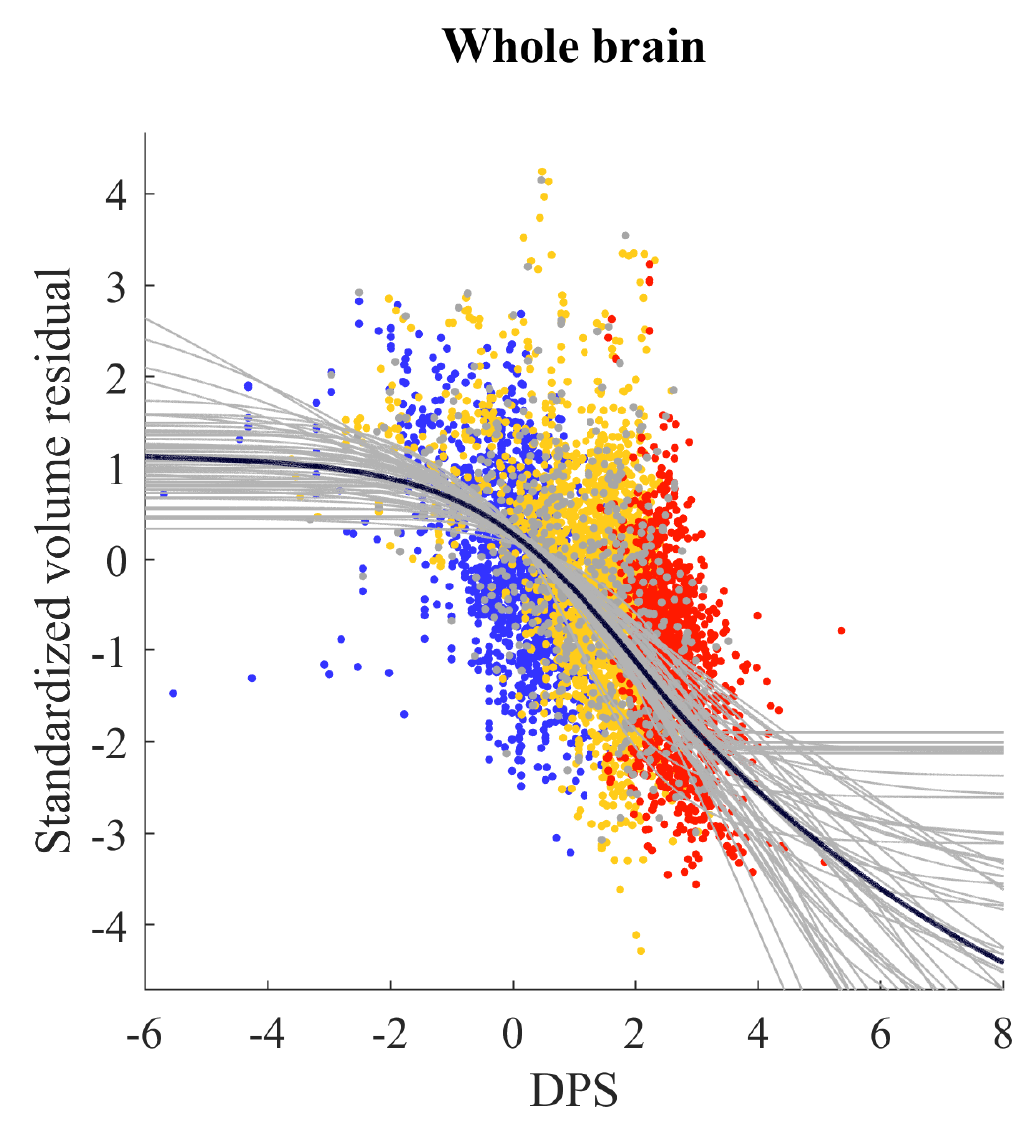}}
\end{subfigure}
\begin{subfigure}[t]{0.221\textwidth}
\raisebox{-\height}{\includegraphics[scale=0.4]{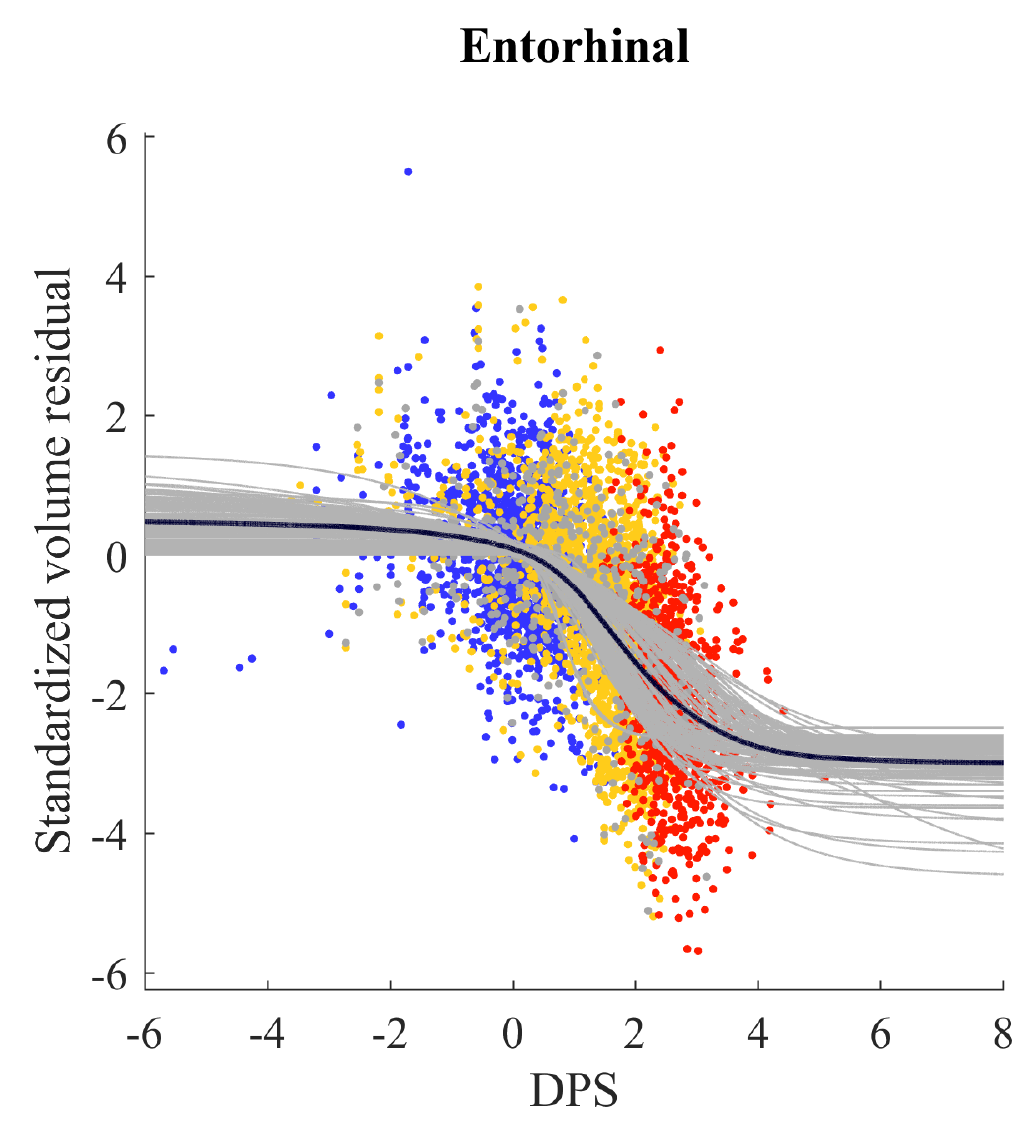}}
\end{subfigure}
\begin{subfigure}[t]{0.321\textwidth}
\raisebox{-\height}{\includegraphics[scale=0.4]{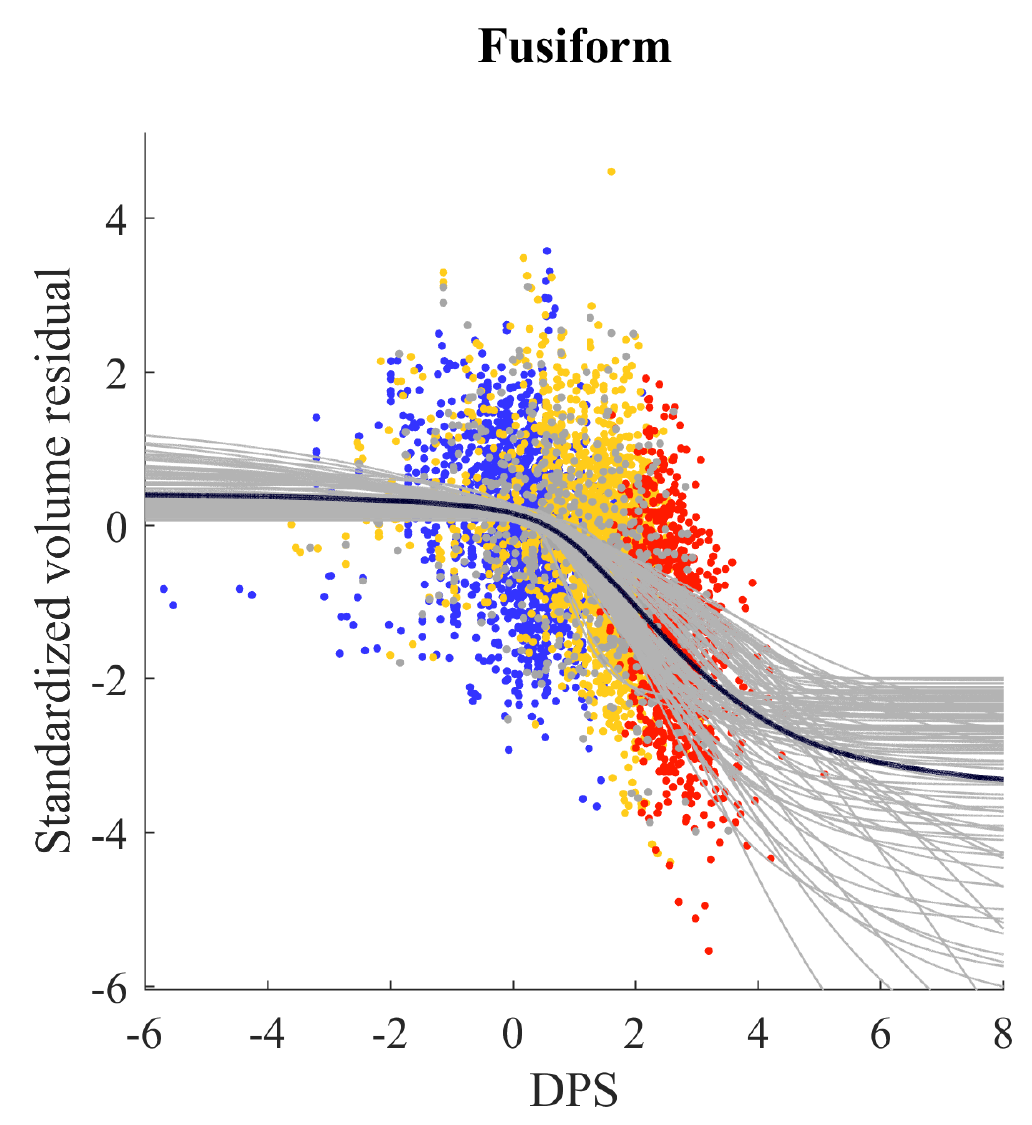}}
\end{subfigure}
\caption{Estimated curves per bootstrap (in gray) for the ADNI biomarkers using the modified Stannard function and the logistic loss. The average of the bootstrapped curves per biomarker is shown as the black curve.}
\label{figure1}
\end{figure*}

To further investigate the stability and robustness of the model with the chosen configuration of logistic and loss functions, we visualize the fitted trajectories for each of the 100-bootstrap runs together with their average per biomarker in Figure \ref{figure1}.

\begin{figure*}[!t]
\centering
\begin{subfigure}[t]{0.42\textwidth}
\raisebox{-\height}{\includegraphics[scale=0.7]{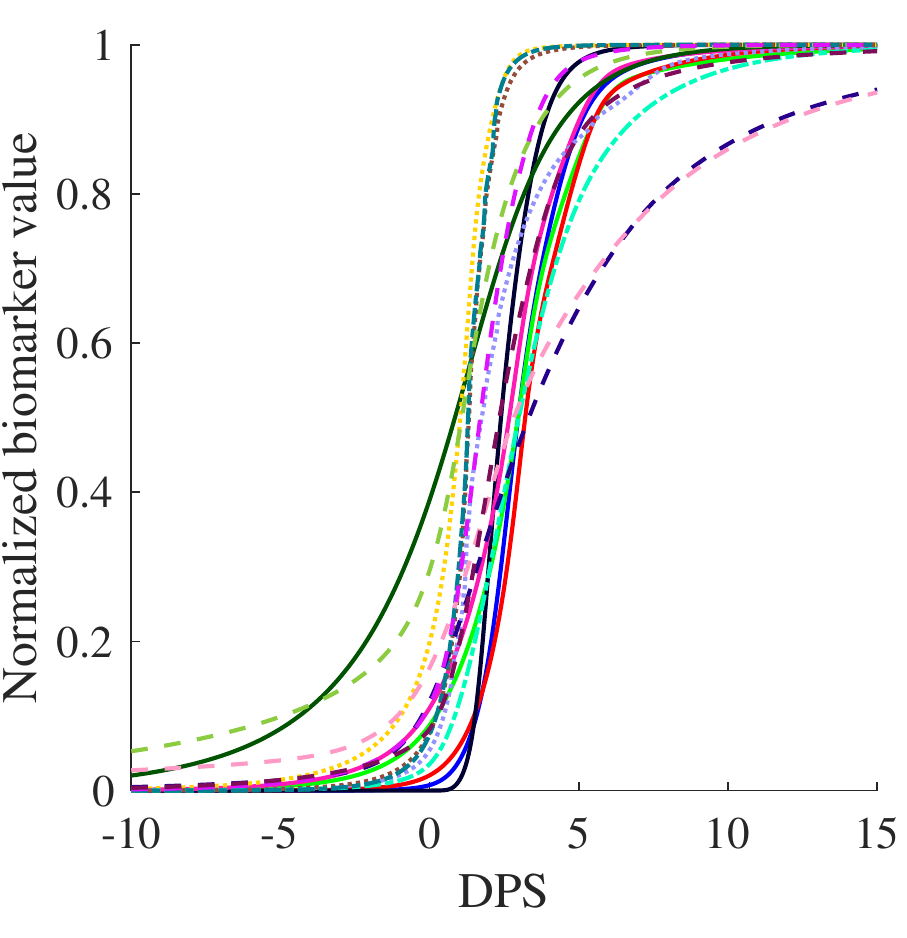}}
\caption{The entire trajectory of all biomarkers.}
\end{subfigure}
\begin{subfigure}[t]{0.49\textwidth}
\raisebox{-\height}{\includegraphics[scale=0.7]{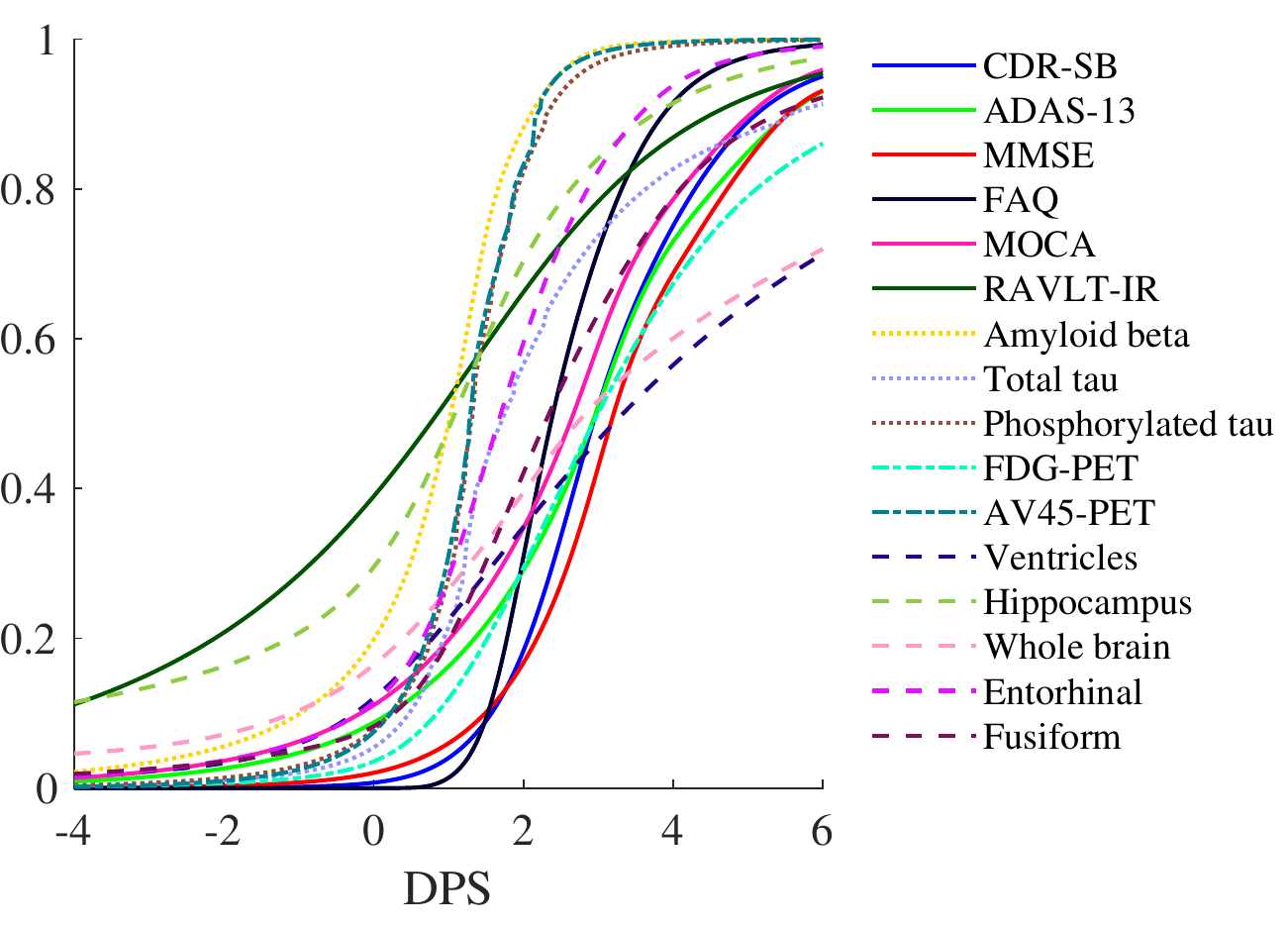}}
\caption{A zoom on the DPS axis showing the most dynamic area.}
\end{subfigure}
\caption{The average of the normalized curves of the ADNI biomarkers across 100 bootstraps.}
\label{figure2}
\end{figure*}

\begin{figure}[!t]
\centering
\includegraphics[scale=0.67]{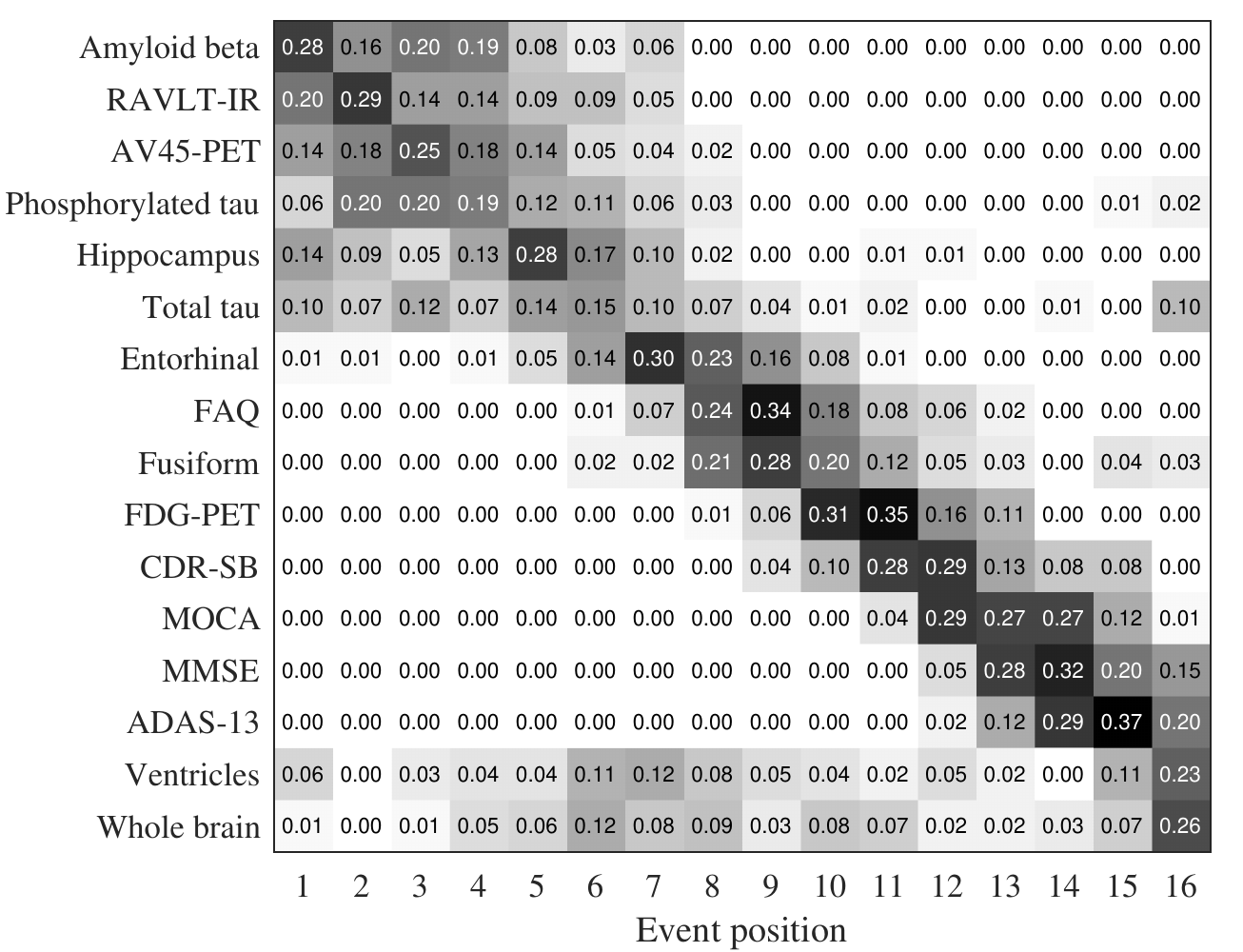}
\caption{Temporal ordering of the ADNI biomarkers in the disease course obtained using inflection points and quantified through 100 bootstraps. The values in the matrix represent the frequency of occurrences (probabilities) and the units in the x-axis indicate the relative ordering of the biomarkers.}
\label{figure3}
\end{figure}

\subsection{Temporal ordering of biomarkers}

To indicate the timing and the dynamics of the different biomarkers relative to each other, Figure \ref{figure2} shows the average curves scaled to $[0, 1]$ using the estimated upper and lower asymptotes per biomarker and superimposed in the same figure. The distribution of the inflection points of the biomarkers, quantified through bootstrapping, can be used to see how biomarkers proceed in the course of AD with respect to each other. The inflection point is considered a turning point at which the direction of biomarker curvature changes. Figure \ref{figure3} displays the temporal ordering of the ADNI biomarkers based on the estimated inflection points. As can be seen, CSF and PET biomarkers, as well as RAVLT-IR, precede all other biomarkers followed by MRI biomarkers and cognitive tests. These findings are in line with the results of \citet{Jedynak2012,Jedynak2015,Young2015,Bilgel2019}. More interestingly, RAVLT-IR starts becoming abnormal early in the disease course which is consistent with several clinical studies concluding that some cognitive tests including RAVLT are significant predictors that can predict neurodegenerative changes up to 10 years before clinical diagnosis \citep{Tierney2005,Landau2010,Zandifar2019}. However, some of the MRI biomarkers such as the ventricles and whole-brain are noisy measurements for modeling the progression of AD in this dataset, as also seen in Figure \ref{figure3}. It is important to note that the inflection points are utilized to order the biomarkers in the disease course. These points do not measure when the biomarkers start becoming abnormal and hence, cannot be used for early abnormality detection.

\subsection{DPS distribution versus biomarker timing}

\begin{figure}[!t]
\centering
\includegraphics[scale=0.64]{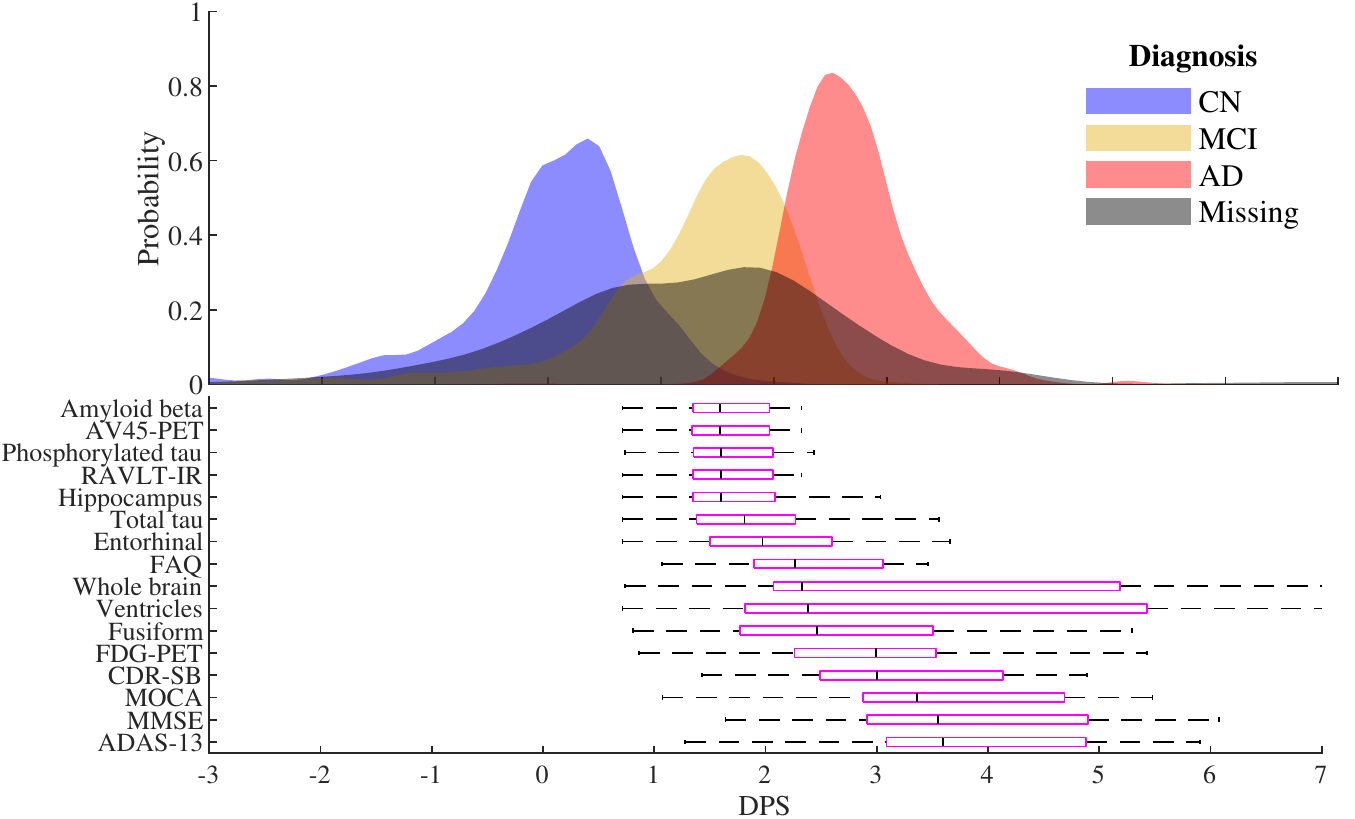}
\caption{Estimated class-conditional likelihoods using the DPSs obtained from 100 ADNI-trained bootstraps. The box plots indicate the 25th to 75th percentiles of the estimated inflection points per biomarker, centrally marked with the median, and they are extended to the most extreme non-outlier inflection points using dashed lines.}
\label{figure4}
\end{figure}

Figure \ref{figure4} shows the variance of the estimated inflection points per biomarker alongside the estimated class-conditional likelihoods of the obtained DPSs from 100 bootstraps. As can be seen, there are moderate overlaps between the DPS distributions of CN-MCI and MCI-AD while the CN and AD groups can be discriminated easily. Moreover, the estimated inflection points per biomarker are almost in line with those of the hypothetical model by \citet{Jack2010} that illustrates when biomarkers are dynamic versus disease stages. Especially, inflection points of the MRI biomarkers (brain structure) are mainly located in the MCI stage while those of the cognitive tests (memory), except for RAVLT-IR, lie on the AD stage.

\begin{table*}[b]
\centering
\normalsize
\caption{Test modeling performance of different methods as NMAE (mean$\pm$SD) for ADNI and NACC biomarkers. Note that ADNI has $16$ biomarkers while NACC has only $6$ biomarkers in common between the two datasets. All the NMAEs are significantly different ($p < 0.001$).}
\label{table6}
\renewcommand{\arraystretch}{1.25}
\centering
\begin{tabular}{lccc}
\toprule
 & \multicolumn{2}{c}{Within cohort} & Across cohort \\
Method & ADNI & NACC & ADNI to NACC \\
\bottomrule
\citet{Jedynak2015} & 1.552$\pm$0.069 & 1.040$\pm$0.210 & 2.665$\pm$0.311 \\
The proposed method & 0.991$\pm$0.023 & 0.833$\pm$0.061 & 1.182$\pm$0.087 \\
\toprule
\end{tabular}
\end{table*}

\subsection{Predicting biomarker values}\label{predict_dps}

The biomarker-specific parameters estimated using the bootstrapped training set are applied to map the ages of test individuals to DPSs using Equation (\ref{subject-specific-eq}). The obtained DPSs are then fed to the estimated biomarker functions in each bootstrap. Table \ref{table6} shows the test NMAEs of the 100-times bootstrapped ADNI dataset for the proposed model and the analogous model by \citet{Jedynak2015} that independently fits the basic sigmoid function using an unconstrained, L2-norm loss function. The proposed model significantly ($p < 0.001$) outperforms the analogous model with an average NMAE of $0.991$ vs. $1.552$ and an average BIC of $1.828 \times 10^4$ vs. $3.303 \times 10^4$. Table \ref{table_list} shows the average test MAE per biomarker across 100 bootstraps.

\begin{figure*}[t]
\centering
\begin{subfigure}[t]{0.221\textwidth}
\raisebox{-\height}{\includegraphics[scale=0.4]{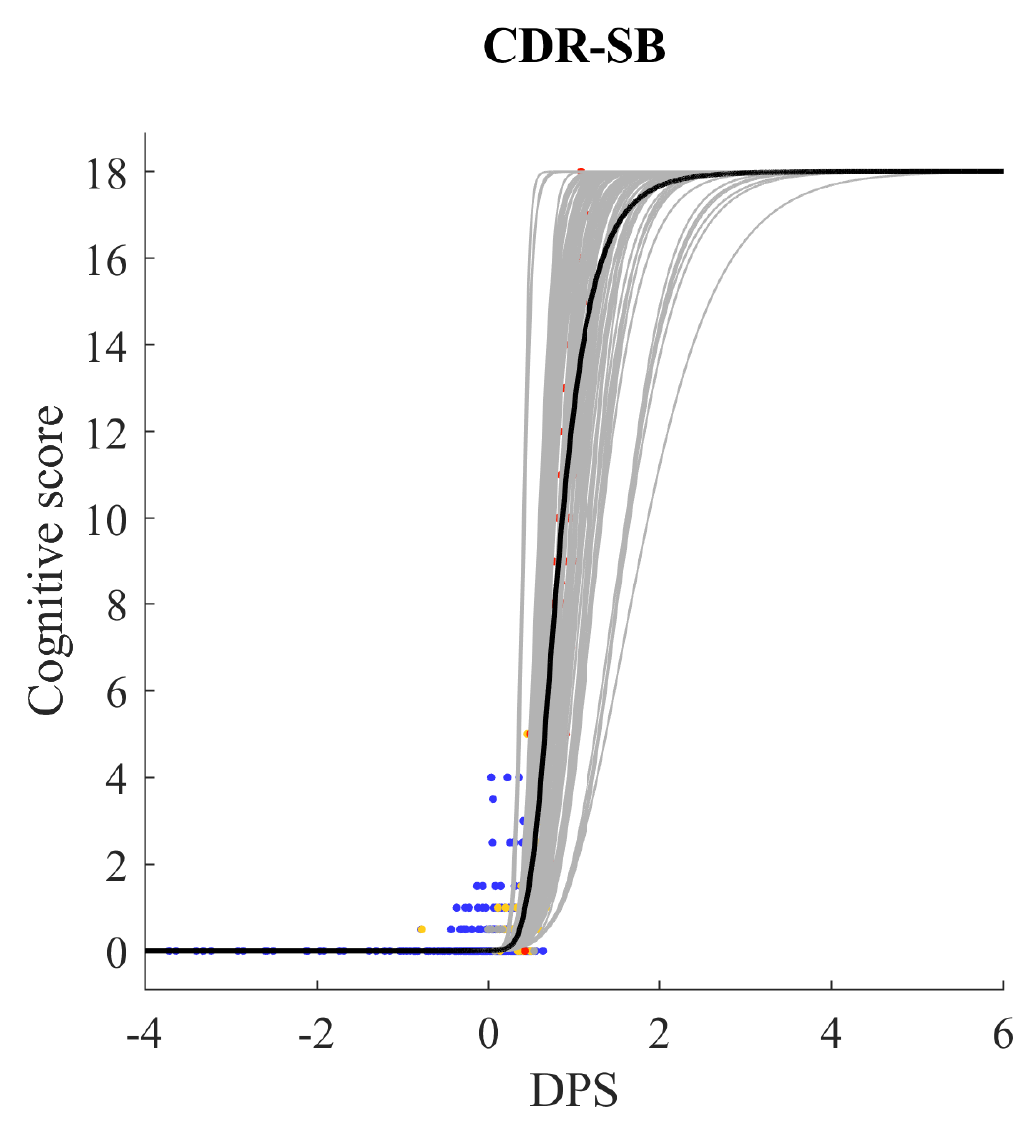}}
\end{subfigure}
\begin{subfigure}[t]{0.221\textwidth}
\raisebox{-\height}{\includegraphics[scale=0.4]{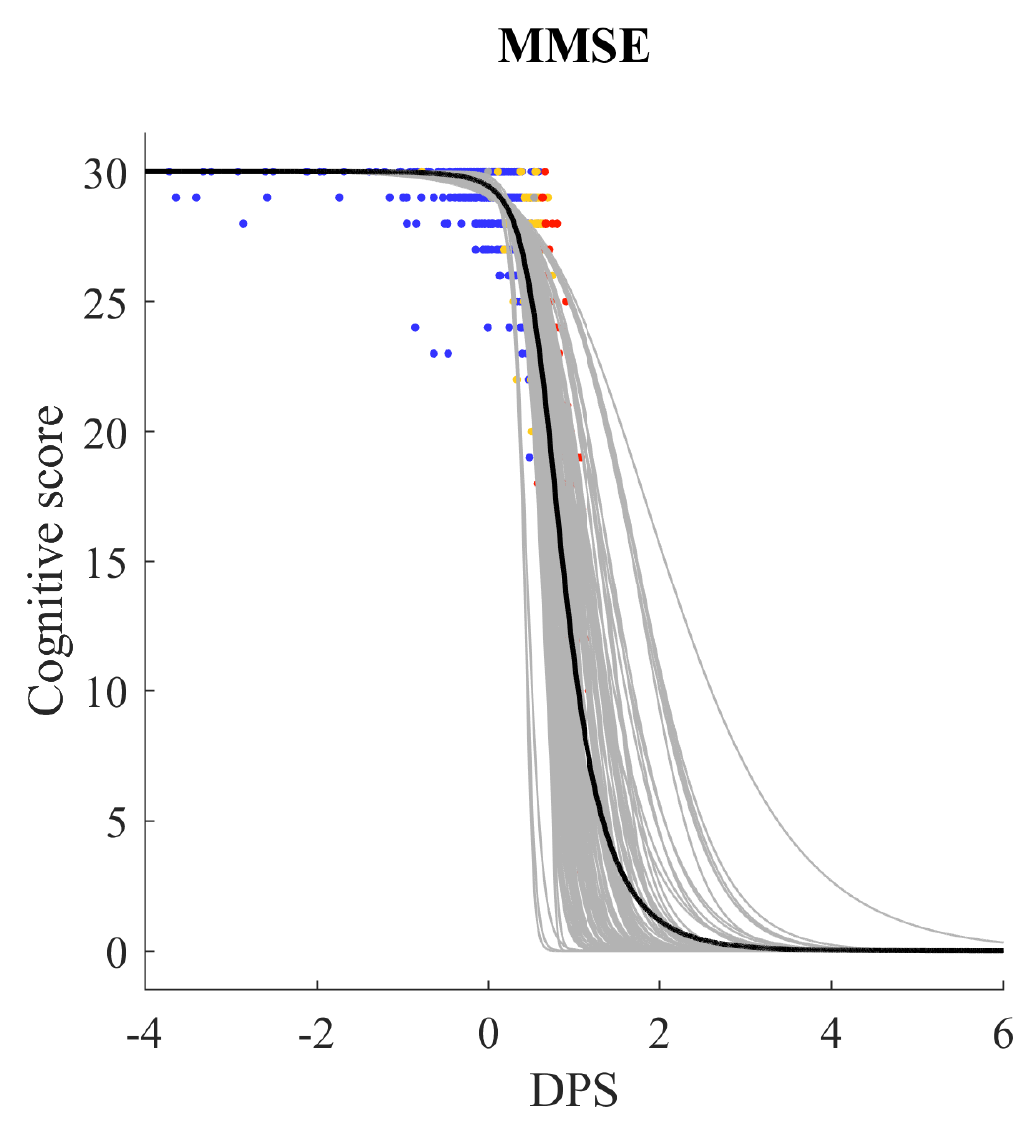}}
\end{subfigure}
\begin{subfigure}[t]{0.221\textwidth}
\raisebox{-\height}{\includegraphics[scale=0.4]{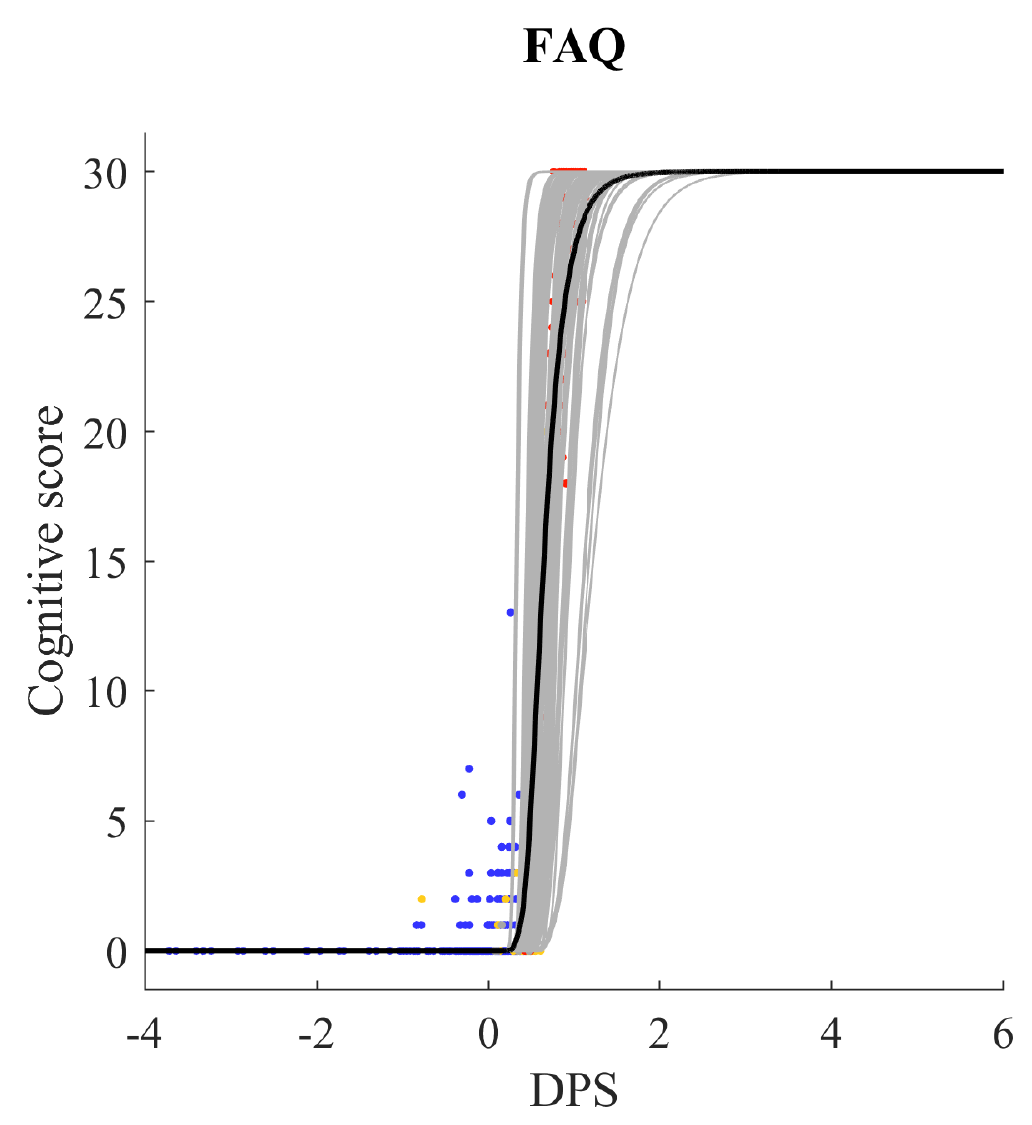}}
\end{subfigure}
\begin{subfigure}[t]{0.321\textwidth}
\raisebox{-\height}{\includegraphics[scale=0.4]{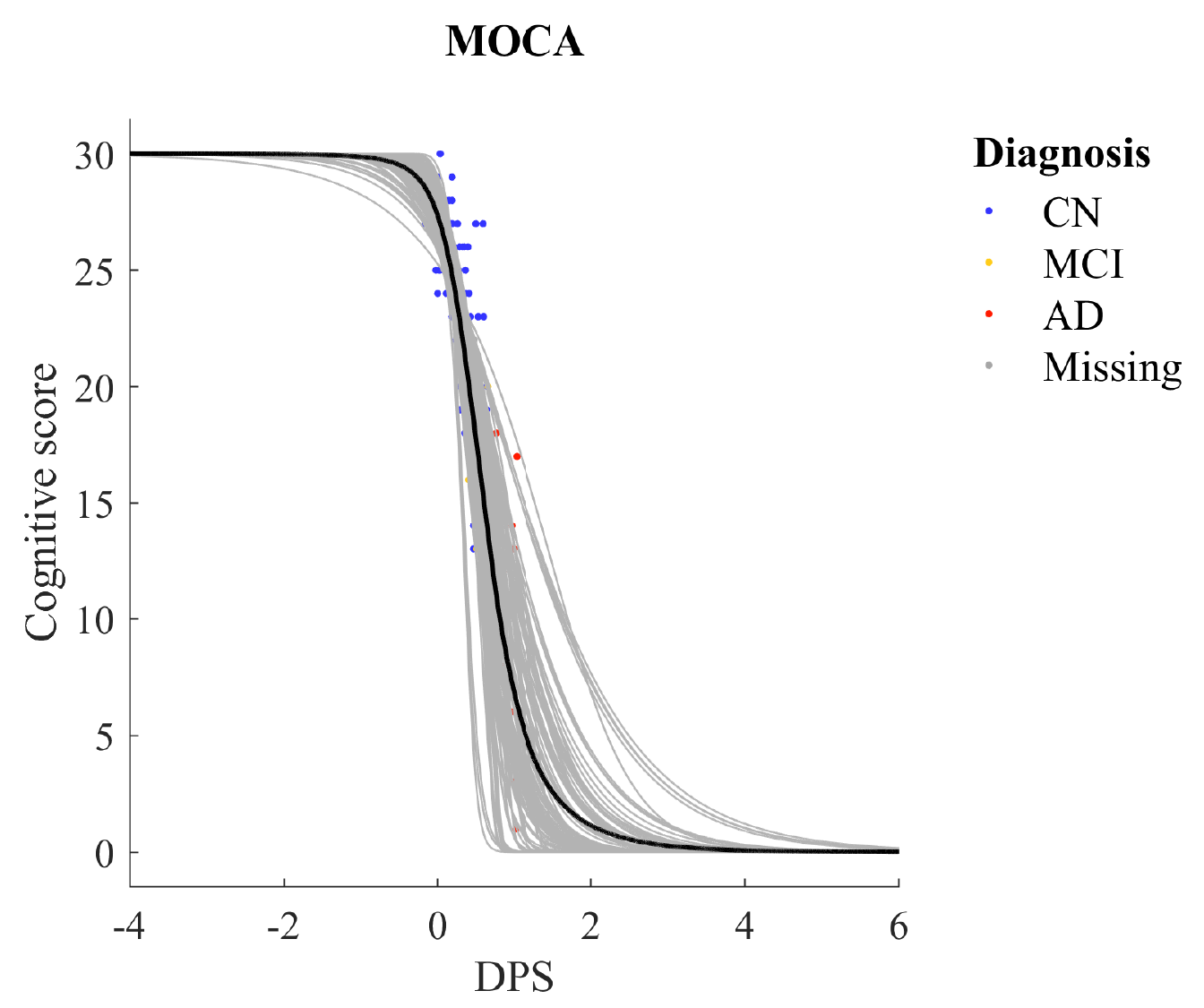}}
\end{subfigure}
\begin{subfigure}[t]{0.221\textwidth}
\raisebox{-\height}{\includegraphics[scale=0.4]{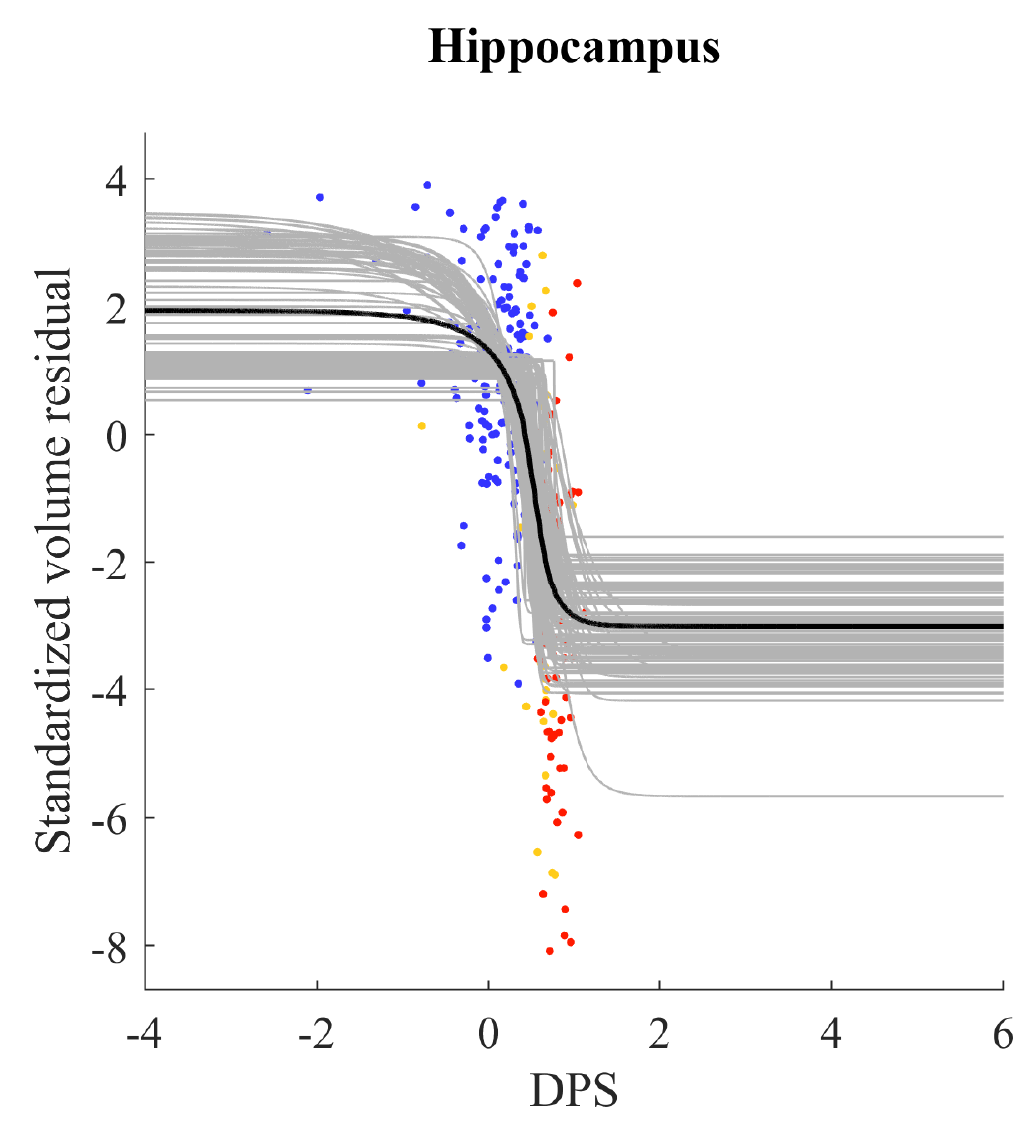}}
\end{subfigure}
\begin{subfigure}[t]{0.221\textwidth}
\raisebox{-\height}{\includegraphics[scale=0.4]{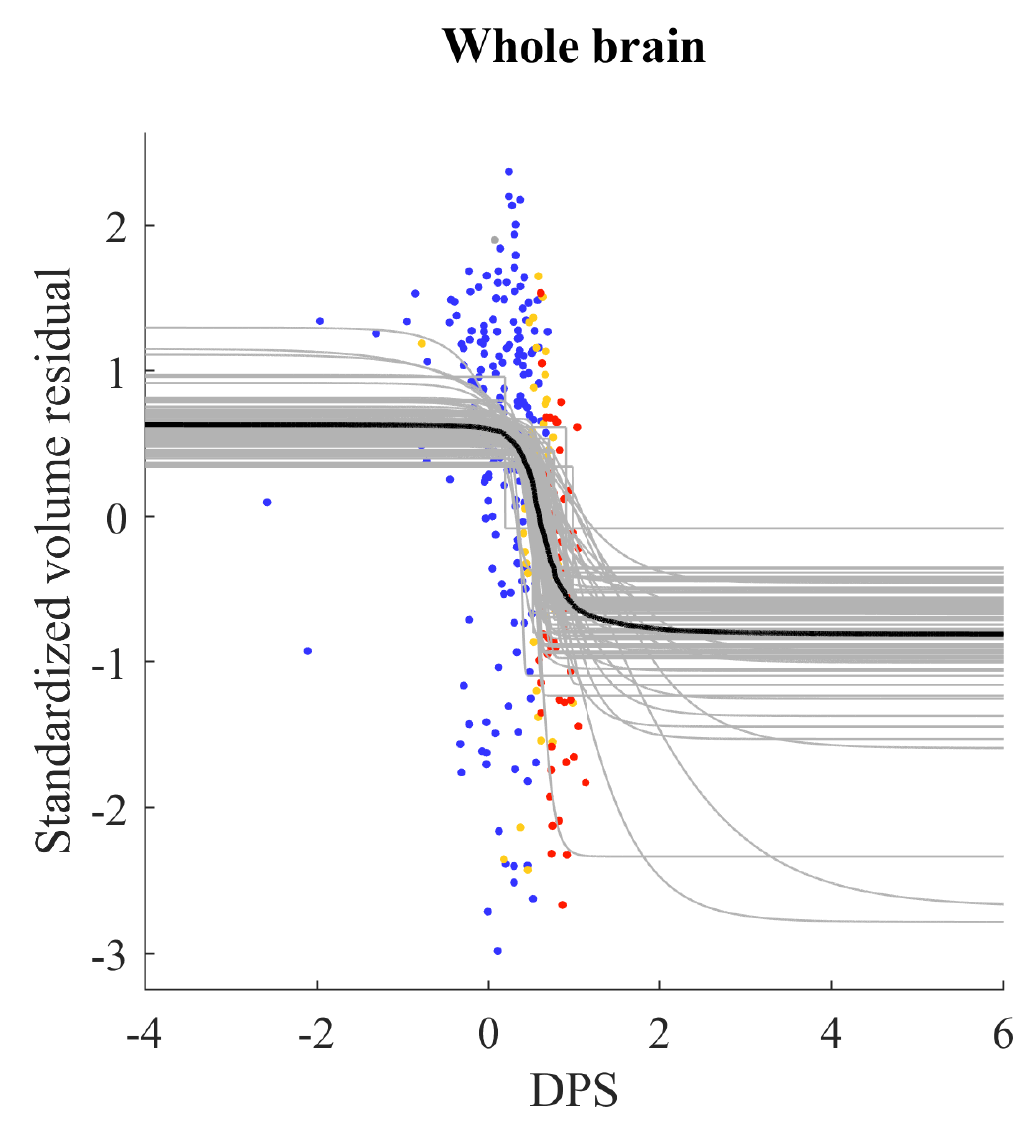}}
\end{subfigure}
\begin{subfigure}[t]{0.221\textwidth}
\raisebox{-\height}{\includegraphics[scale=0.4]{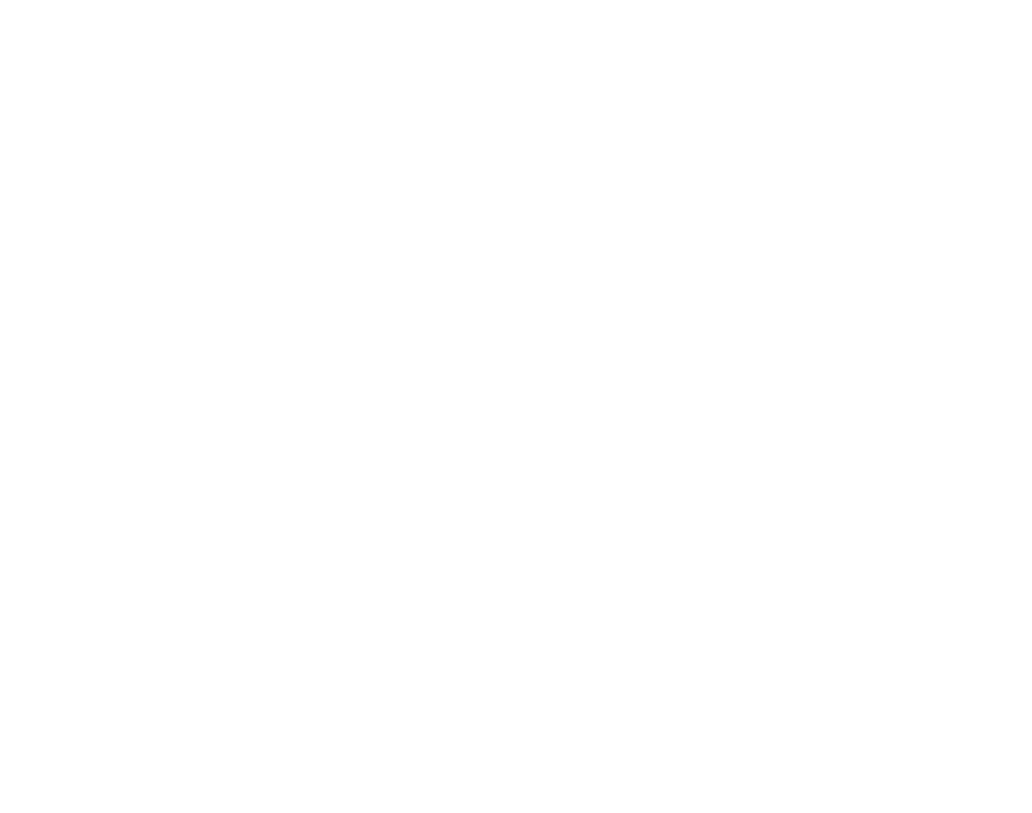}}
\end{subfigure}
\begin{subfigure}[t]{0.32\textwidth}
\raisebox{-\height}{\includegraphics[scale=0.4]{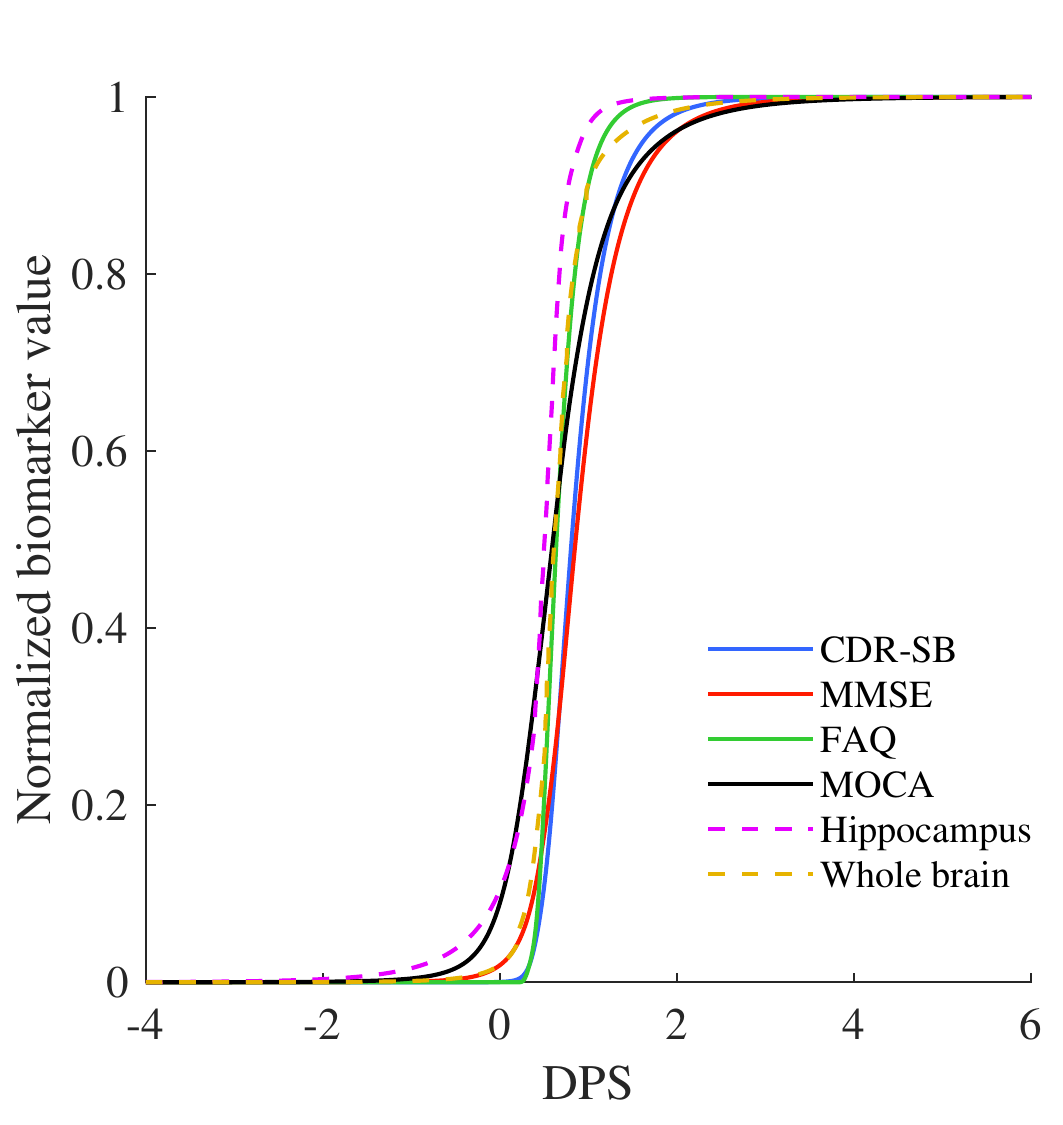}}
\end{subfigure}
\caption{Estimated curves per bootstrap (in gray) for the NACC biomarkers using the modified Stannard function and the logistic loss. The average of the bootstrapped curves per biomarker is shown as the black curve. The last subfigure shows the average of the normalized curves of the NACC biomarkers across 100 bootstraps.}
\label{figure5}
\end{figure*}

\begin{table*}[!h]
\centering
\normalsize
\caption{Detailed information about the utilized ADNI biomarkers.}
\label{table_list}
\renewcommand{\arraystretch}{1.25}
\centering
\begin{tabular}{p{3cm} p{5cm} p{2.5cm} c c c}
\toprule
Biomarker & Interpretation & Unit & Range & Inflection point & Test MAE \\
 &  &  &  & (median) & (mean) \\
\bottomrule
CDR-SB & The sum of scores of six sets of questions. Lower values indicate less cognitive dysfunction. & Cognitive score & [0, 18] & 3.003 & 0.562 \\
ADAS-13 & The sum of scores of 13 itemized tasks. Lower values indicate less cognitive dysfunction. & Cognitive score & [0, 85] & 3.596 & 4.236 \\
MMSE & The sum of scores of a set of questions. Lower values indicate more cognitive dysfunction. & Cognitive score & [0, 30] & 3.552 & 1.506 \\
FAQ & The sum of scores of 10 sets of questions. Lower values indicate less cognitive dysfunction. & Cognitive score & [0, 30] & 2.264 & 1.415 \\
MOCA & The sum of scores of 30 questions. Lower values indicate more cognitive dysfunction. & Cognitive score & [0, 30] & 3.363 & 2.154 \\
RAVLT-IR & The sum of scores from five trials in remembering a list of 15 words immediately after each trial. Lower values indicate more cognitive dysfunction. & Cognitive score & [0, 75] & 1.600 & 5.983 \\
CSF amyloid-beta & The concentration level of brain beta-amyloid protein. Lower values indicate more concentration. & Picograms per milliliter & (0, $\infty$) & 1.591 & 374.4 \\
CSF total tau and phosphorylated tau & The concentration level of neurofibrillary tangles of brain tau protein. Lower values indicate less concentration. & Picograms per milliliter & (0, $\infty$) & \parbox[t]{1cm}{1.811 \\ 1.600} & \parbox[t]{1cm}{95.19 \\ 10.10} \\
FDG-PET & The regional cerebral metabolic rate of glucose. Lower values indicate less activity. & Standardized uptake value ratio & (0, $\infty$) & 2.995 & 0.104 \\
AV45-PET & The cerebral amyloid deposition. Lower values indicate less deposition. & Standardized uptake value ratio & (0, $\infty$) & 1.591 & 0.151 \\
Adjusted T1-weighted brain MRI volumes of ventricles, hippocampus, whole brain, fusiform, and entorhinal cortex & The regional brain atrophies. Except in the case of ventricles, lower values indicate more atrophy. & Standardized volume residual & ($-\infty$, $\infty$) & \parbox[t]{1cm}{2.385 \\ 1.600 \\ 2.328 \\ 1.973 \\ 2.461} & \parbox[t]{1cm}{0.899 \\ 0.791 \\ 0.716 \\ 0.883 \\ 0.789} \\
\toprule
\end{tabular}
\end{table*}

\subsection{classifying clinical status}

To evaluate the diagnostic predictive performance, the obtained training DPSs are used to estimate class-conditional likelihood functions per bootstrap using KDE and fed to a three-class Bayesian classifier with prior probabilities proportional to the number of training observations in each class. The classifiers, one for each bootstrap, are applied to the test DPSs estimated as described in Section \ref{predict_dps} to compute the posterior probabilities of the clinical labels. The proposed model achieves an AUC of $0.931\pm0.004$ in classifying the clinical status of the test ADNI subjects per visit, which reveals the effect of modeling on classification performance.

The obtained posterior probabilities from the different classifiers can be combined using ensemble learning techniques to potentially improve prediction performance and robustness \citep{Kuncheva2014}. For example, by fusing the posteriors based on taking the average of the within-class posteriors over an ensemble of models from different bootstraps (bagging), the AUC of the proposed method increases to $0.934$.

\subsection{Comparison with state-of-the-art results}

In order to fairly compare our results with those of state-of-the-art methods, we apply the proposed method to the TADPOLE training and test subsets of D1 and D2 using the same 16 ADNI biomarkers. The proposed model achieves an average AUC of $0.937$ which is on a par with the best performance of TADPOLE with an average AUC of $0.931$ \citep{Marinescu2020}. Besides, our obtained average MAE of $3.93$ for ADAS-13 outperforms the best reported result with an average MAE of $4.70$. However, the proposed model does not perform well on the normalized ventricles compared to the best reported result with an average MAE of $0.0086$ vs.$0.0041$.

Next, we employ the same ADNI data splits and biomarkers as used by \citet{Bilgel2019} and make a head-to-head comparison with the results reported in the aforementioned study. This also enables a head-to-head comparison with both \citet{Li2019} and \citet{Lorenzi2017} based on their results reported by \citet{Bilgel2019}. To do so, biomarker trajectories need to be described as a function of time-from-AD-conversion. Hence, inspired by \citet{Bilgel2019}, we select any subjects converting to AD and calculate the time from AD conversion using the difference between the visiting age and the age at which the first AD status is diagnosed. The corresponding DPSs are then mapped to the obtained times from the AD conversion of the selected subjects using a linear regression model. These estimates can later be used to calculate the time-from-AD-conversion for any subject’s visits using the estimated DPSs. Since the time-from-AD-conversion is a linear function of DPS, i.e., $\hat{m}_0 + \hat{m}_1 s_{i,j}$, we can adjust the biomarker parameters as $b_k = b_k / \hat{m_1}$ and $c_k = \hat{m}_0 + \hat{m}_1 c_k$ to obtain biomarker trajectories as a function of time-from-AD-conversion. The obtained results indicate that the proposed model outperforms \citet{Bilgel2019} with a root-mean-square-error of $0.68$ vs. $1.48$; yet it has a larger maximum absolute error ($4.20$ vs. $3.79$).

\subsection{Generalizability across cohorts}

As the final set of experiments, the generalizability of the proposed model to an independent cohort is assessed using the NACC data. First, the same configuration of logistic function and M-estimator, i.e., the modified Stannard and logistic loss is applied to model the progression of AD within NACC. Figure \ref{figure5} depicts the modeled NACC biomarkers for 100 bootstraps. Second, the optimal model previously trained on ADNI is utilized to predict the NACC test measurements using the estimates of the common ADNI-NACC biomarkers, i.e., CDR-SB, MMSE, FAQ, MOCA, hippocampus, and whole brain. Table \ref{table6} compares the modeling performance of the ADNI-trained and NACC-trained models applied to the NACC test set. As it can be noticed from the obtained results, the previously selected configuration for training ADNI data is also a good choice when applied to NACC data. Moreover, the proposed model significantly ($p < 0.001$) outperforms the analogous model of \citet{Jedynak2015} in all cases. Additionally, modeling performance of the proposed method degrades less than that of the analogous model of \citet{Jedynak2015} when applying the ADNI-trained model to the NACC test set, which indicates the generalizability of the proposed method across cohorts. It should also be noted that the utilized NACC subset have fewer biomarkers and measurements than the used ADNI subset, which likely is the reason why it results in a smaller within-cohort modeling error.

We also apply the ADNI and NACC trained classifiers to the estimated test NACC DPSs to classify the clinical status per subject per visit. The proposed method achieves AUCs of $0.929\pm0.012$ and $0.928\pm0.016$, respectively. This reveals that diagnostic performance improves when applying the ADNI-trained model to the NACC test set.

\section{Conclusions}

In this paper, a robust parametric model of Alzheimer's disease progression was proposed based on alternating M-estimation using the logistic loss to address potential curve-fitting problems such as outliers. The proposed method linearly transformed individuals' ages to disease progression scores and jointly fitted modified Stannard functions to the longitudinal dynamics of biomarkers. The estimated parameters were then used to temporally order the biomarkers in the disease course and to predict biomarker values as well as to classify the clinical status per subject visit in an independent test set. The obtained results showed the superiority of the proposed method over the state-of-the-art results in terms of prediction performance, and this method generalizes well across cohorts.

The proposed approach can be applied to different time-series data including missing data points and labels, or to biomarkers with other characteristics than the monotonic behavior that one typically encounters in, for example, neurodegenerative disease progression modeling using MRI/PET biomarkers, as long as suitable functions are used for biomarker modeling. Moreover, as an alternative to using M-estimators, resistant estimators such as the least trimmed sum of squares and least median of squares \citep{Rousseeuw1987} with higher breakdown points can be used to fit biomarker trajectories. Though, this will result in an additional parameter to be optimized for the coverage (range) needed for trimming the residuals.


\section*{Disclosures}

M. Nielsen is a shareholder in Biomediq A/S and Cerebriu A/S. A. Pai is a shareholder in Cerebriu A/S. The remaining authors report no disclosures.

\section*{Acknowledgments}

This project has received funding from the European Union's Horizon 2020 research and innovation programme under the Marie Sk{\l}odowska-Curie grant agreement No 721820.

\noindent \textbf{\textit{ADNI acknowledgments--}}Data collection and sharing for this project was funded by the Alzheimer's Disease Neuroimaging Initiative (ADNI) (National Institutes of Health Grant U01 AG024904) and DOD ADNI (Department of Defense award number W81XWH-12-2-0012). ADNI is funded by the National Institute on Aging, the National Institute of Biomedical Imaging and Bioengineering, and through generous contributions from the following: AbbVie, Alzheimer's Association; Alzheimer's Drug Discovery Foundation; Araclon Biotech; BioClinica, Inc.; Biogen; Bristol-Myers Squibb Company; CereSpir, Inc.; Cogstate; Eisai Inc.; Elan Pharmaceuticals, Inc.; Eli Lilly and Company; EuroImmun; F. Hoffmann-La Roche Ltd. and its affiliated company Genentech, Inc.; Fujirebio; GE Healthcare; IXICO Ltd.; Janssen Alzheimer Immunotherapy Research \& Development, LLC.; Johnson \& Johnson Pharmaceutical Research \& Development LLC.; Lumosity; Lundbeck; Merck \& Co., Inc.; Meso Scale Diagnostics, LLC.; NeuroRx Research; Neurotrack Technologies; Novartis Pharmaceuticals Corporation; Pfizer Inc.; Piramal Imaging; Servier; Takeda Pharmaceutical Company; and Transition Therapeutics. The Canadian Institutes of Health Research is providing funds to support ADNI clinical sites in Canada. Private sector contributions are facilitated by the Foundation for the National Institutes of Health (www.fnih.org). The grantee organization is the Northern California Institute for Research and Education, and the study is coordinated by the Alzheimer's Therapeutic Research Institute at the University of Southern California. ADNI data are disseminated by the Laboratory for Neuro Imaging at the University of Southern California.

\noindent \textbf{\textit{TADPOLE acknowledgments--}}This work uses the TADPOLE data sets (https://tadpole.grand-challenge.org) constructed by the EuroPOND consortium (http://europond.eu) funded by the European Union's Horizon 2020 research and innovation programme under grant agreement No 666992.

\noindent \textbf{\textit{NACC acknowledgments--}}The NACC database is funded by NIA/NIH Grant U01 AG016976. NACC data are contributed by the NIAfunded ADCs: P30 AG019610 (PI Eric Reiman, MD), P30 AG013846 (PI Neil Kowall, MD), P50 AG008702 (PI Scott Small, MD), P50 AG025688 (PI Allan Levey, MD, PhD), P50 AG047266 (PI Todd Golde, MD, PhD), P30 AG010133 (PI Andrew Saykin, PsyD), P50 AG005146 (PI Marilyn Albert, PhD), P50 AG005134 (PI Bradley Hyman, MD, PhD), P50 AG016574 (PI Ronald Petersen, MD, PhD), P50 AG005138 (PI Mary Sano, PhD), P30 AG008051 (PI Thomas Wisniewski, MD), P30 AG013854 (PI M. Marsel Mesulam, MD), P30 AG008017 (PI Jeffrey Kaye, MD), P30 AG010161 (PI David Bennett, MD), P50 AG047366 (PI Victor Henderson, MD, MS), P30 AG010129 (PI Charles DeCarli, MD), P50 AG016573 (PI Frank LaFerla, PhD), P50 AG005131 (PI James Brewer, MD, PhD), P50 AG023501 (PI Bruce Miller, MD), P30 AG035982 (PI Russell Swerdlow, MD), P30 AG028383 (PI Linda Van Eldik, PhD), P30 AG053760 (PI Henry Paulson, MD, PhD), P30 AG010124 (PI John Trojanowski, MD, PhD), P50 AG005133 (PI Oscar Lopez, MD), P50 AG005142 (PI Helena Chui, MD), P30 AG012300 (PI Roger Rosenberg, MD), P30 AG049638 (PI Suzanne Craft, PhD), P50 AG005136 (PI Thomas Grabowski, MD), P50 AG033514 (PI Sanjay Asthana, MD, FRCP), P50 AG005681 (PI John Morris, MD), P50 AG047270 (PI Stephen Strittmatter, MD, PhD).

\bibliographystyle{model2-names}
\bibliography{refs}






\end{document}